\documentclass[aps,prl,reprint,superscriptaddress]{revtex4-2}
\usepackage{amsmath,amssymb,bm,graphicx}
\usepackage{microtype}
\usepackage[hidelinks]{hyperref}

\newcommand{\dd}{\mathrm{d}}
\newcommand{\ii}{\mathrm{i}}
\newcommand{\e}{\mathrm{e}}
\newcommand{\Tr}{\mathrm{Tr}}
\newcommand{\tr}{\mathrm{tr}}
\newcommand{\BZ}{\mathrm{BZ}}
\newcommand{\Ds}{D_{\rm s}}
\newcommand{\Torb}{T_{\rm orb}}
\newcommand{\TLVP}{T_{\rm LVP}}
\newcommand{\TLVPcl}{T_{\rm LVP}^{\rm cl}}
\newcommand{\Ecal}{\mathcal E}
\newcommand{\Vcal}{\mathcal V}

\begin{document}

\title{Vortexable Chern Bands Sit Exactly on the Topological Floor of Flat-Band Superfluid Stiffness}

\author{Zhi-Wei Wang}
\email{zhiweiwang.phy@gmail.com}
\affiliation{College of Physics, Jilin University, Changchun, 130012, People's Republic of China}
\affiliation{Department of Computer Science, University of York, York YO10 5GH, United Kingdom}

\author{Samuel L. Braunstein}
\email{sam.braunstein@york.ac.uk}
\affiliation{Department of Computer Science, University of York, York YO10 5GH, United Kingdom}

\date{\today}

\begin{abstract}
Vortexable Chern bands host ideal fractional Chern insulators. Paired with time-reversed partners by orbital-diagonal on-site attraction, they are the least stiff flat-band superconductors at fixed Chern number $C$, filling, and pairing amplitude. For uniform pairing and a nondegenerate paired state, the exact $T=0$ stiffness determinant is the topological floor times $1+(\Torb/2\pi|C|)^2\le1+(\TLVP/2\pi|C|)^2$, where $\Torb$ and $\TLVP$ measure departures from vortexability. Enhancement costs vortexability, and the inequality is strict only if internal components share a site, as in heterostrained bilayers.
\end{abstract}

\maketitle

Fractional Chern insulators now appear at zero magnetic field in twisted MoTe$_2$ and rhombohedral graphene \cite{Cai2023,Zeng2023,Park2023,XuPRX2023,Lu2024}, and closely related devices superconduct \cite{Han2025,Choi2025}; in twisted MoTe$_2$, superconductivity has been reported in the Chern band that hosts the fractional states \cite{Xu2025}.  Both phases are controlled by the quantum geometry of flat Chern bands.  Vortexable bands, which are ideal in a suitable real-space frame, share the holomorphic structure of the lowest Landau level, and short-range repulsion then has exact fractional Chern insulator ground states \cite{Roy2014,Claassen2015,MeraOzawa2021,Wang2021,LVP2023}.  Flat-band superconductors draw their phase stiffness from the quantum metric, which the Chern number bounds from below \cite{PT2015,Liang2017,Xie2020,HAPSHB2022,Huhtinen2022,Torma2022}.  Ledwith, Vishwanath, and Parker (LVP) asked in their conclusions whether vortexable bands of equal and opposite Chern numbers are ``ideal for other phases such as superconductors'' \cite{LVP2023}.

For the zero-temperature superfluid stiffness we answer this question.  Paired with their time-reversed partners by orbital-diagonal on-site attraction and with uniform pairing, vortexable bands are the least stiff flat-band superconductors at fixed Chern number, filling, and pairing amplitude, in the sense of the geometric mean of the principal stiffnesses.  Their stiffness sits exactly on the topological floor, and any enhancement must be paid for in vortexability.  The converse fails only when internal components share a site or sites would have to collide; we give an exact lattice example and compute the underlying splitting in a physical band.

The answer comes from comparing two optimizations.  The quantum metric of a tight-binding band depends on where the orbitals are placed; the superfluid weight does not.  For a time-reversed pair of isolated flat bands with uniform pairing it is set by the minimal quantum metric \cite{Huhtinen2022}, reached at the natural positions of the orbitals \cite{TP2024}.  LVP's general vortexability measure also minimizes, over coordinate frames, which, as LVP note, had been interpreted for tight-binding models as unit-cell embeddings \cite{LVP2023,SimonRudner2020,Huhtinen2022}.  We show that the two are one least-squares problem on nested spaces, because Cooper pairs relax the phase of every orbital independently, whereas a vortex coordinate moves all orbitals at one point together.  Their difference is a positive-semidefinite splitting defect $\Sigma$.  Recent proofs \cite{HKH2026,WLY2026}, which confirm a conjecture based on bootstrap bounds \cite{GHK2026}, show that the mean-field stiffness of such models is exact when the paired ground state is nondegenerate; our identity is then a statement about the exact stiffness of the band-projected model.

\paragraph{Embeddings and frames.}
Consider a translation-invariant tight-binding model with $N$ point orbitals per unit cell; orbitals at a common position form a \emph{site}, and $N_{\rm s}$ is the number of sites.  We set $\hbar$, the electron charge, and the unit-cell area to one, and $\dd^2k$ carries no factor $(2\pi)^{-2}$.  For an isolated band with Bloch vector $u(\bm k)$ and projector $P$, the quantum geometric tensor $Q_{ij}=\Tr[P\,\partial_iP\,\partial_jP]=g_{ij}-\frac{\ii}{2}\Omega_{ij}$, with $\partial_i=\partial/\partial k_i$, quantum metric $g$, and Berry curvature $\Omega$, integrates to $\bar Q=\int_\BZ Q\,\dd^2k$, whose real part $G$ obeys
\begin{equation}
\det\bar Q=\det G-\pi^2C^2 ,
\label{eq:detQ}
\end{equation}
with $C$ the Chern number.  An embedding $e$ displaces orbital $\alpha$ by $\bm\eta_\alpha$ and replaces $u_\alpha$ by $\e^{-\ii\bm k\cdot\bm\eta_\alpha}u_\alpha$; it changes $G$, but not $C$ \cite{SimonRudner2020,Huhtinen2022}.  We write $\Ecal\simeq\mathbb R^{2N}$ for all embeddings and $\Vcal\subset\Ecal$ for the site-rigid ones, in which the orbitals of each site move together.

LVP's measure is the minimum over lattice-periodic, orientation-preserving diffeomorphisms $\phi$ of the plane, $\TLVP=\min_\phi2\sqrt{\det\bar Q_\phi}$ [their Eqs.~(35) and (39)], and a band is vortexable, with a vortex function of lattice-periodic gradient, if and only if $\TLVP=0$ \cite{LVP2023}.  We transfer this measure to tight-binding models by treating orbitals as points. A point orbital samples $\phi$ only at its own position, so every $\phi$ acts as a site-rigid embedding; conversely, isotopy extension reaches every collision-free site-rigid embedding (End Matter, compare Sec.~VI of Ref.~\cite{WRS2026}).  The LVP orbit is therefore dense in $\Vcal$, and we use the closure value $\TLVPcl=2\min_{e\in\Vcal}\sqrt{\det\bar Q_e}$, the infimum over frames, which is attained by a frame whenever the minimizing site configuration is collision free. In every example below that configuration is collision free, so $\TLVPcl=\TLVP$.  Lifting the site constraint defines the orbital-resolved measure $\Torb=2\min_{e\in\Ecal}\sqrt{\det\bar Q_e}\le\TLVPcl$.  Positions are physical data, since one Bloch Hamiltonian with its orbitals on one site or on two describes two different systems, with the same stiffness and, as we show, different vortexability (compare the embedding dependence of ideality noted in Refs.~\cite{Zhao2026,WRS2026}).

\begin{figure}[t]
\includegraphics[width=\columnwidth]{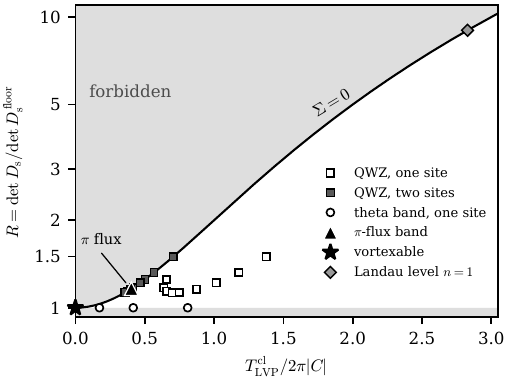}
\caption{Trade-off plane.  Each band, in a given physical realization, is a point with $R=\det\Ds/\det\Ds^{\rm floor}=1+(\Torb/2\pi|C|)^2$ and abscissa $\TLVPcl/2\pi|C|$.  The theorem forbids the upper shaded region; on its boundary curve $\Sigma=0$.  The Chern floor excludes $R<1$ (lower shaded strip).  Squares show the QWZ band of Fig.~\ref{fig:qwz}, $-1.8\le M\le-0.6$, with its orbitals on two sites (filled) or one site (open).  Circles show the theta band on one site, $d=0.1$, $0.25$, $0.5$ (Fig.~\ref{fig:theta}).  The star marks vortexable bands, among them the theta band on two sites and the honeycomb band of Ref.~\cite{LiWu2026}.  The triangle marks the $\pi$-flux band \cite{HBC2020,GHK2026}.  The diamond marks the time-reversed $n=1$ Landau level (continuum, mean field).  Heterostrained twisted bilayer graphene, at abscissa $2.5\times10^{-3}$ and $\det G_\star/\pi^2C^2-1<3\times10^{-8}$ (a geometric ordinate, since its density is not uniform), is not resolved.}
\label{fig:tradeoff}
\end{figure}

\paragraph{Gram identity and theorem.}
With $w_\alpha=|u_\alpha|^2$, let $W_{\alpha\beta}=\int_\BZ(w_\alpha\delta_{\alpha\beta}-w_\alpha w_\beta)\,\dd^2k\succeq0$ and $b_{\alpha j}=-\int_\BZ\mathrm{Im}[u_\alpha^*((1-P)\partial_ju)_\alpha]\,\dd^2k$.  The integrated metric is quadratic in the displacements \cite{TP2024,WRS2026}.  Because no metric component can become negative, every column of $\bm b$ lies in the range of $W$ (End Matter), and completing the square about $\bm\eta^\star=-W^+\bm b$, with $W^+$ the pseudoinverse, gives
\begin{equation}
G_e=G_\star+\Gamma(\bm\eta-\bm\eta^\star),\quad \Gamma(\bm\zeta)_{ij}=\textstyle\sum_{\alpha\beta}\zeta_{\alpha i}W_{\alpha\beta}\zeta_{\beta j}\succeq0 .
\label{eq:gram}
\end{equation}
The natural embedding $\bm\eta^\star$ is therefore minimal in the Loewner order.  This minimality, uniqueness up to translation apart from trivially removable orbitals, and invariance under symmetries that permute orbitals up to phases are in Ref.~\cite{WRS2026}; the minimal metric, and its attainment at high-symmetry positions, are in Ref.~\cite{Huhtinen2022}.  Minimizing over $\Vcal$ instead gives
\begin{equation}
G_\Vcal=G_\star+\Sigma,\qquad\Sigma=\Gamma(\bm\eta^\Vcal-\bm\eta^\star)\succeq0,
\label{eq:sigma}
\end{equation}
with $\bm\eta^\Vcal$ the $W$-orthogonal projection of $\bm\eta^\star$ onto $\Vcal$.  The splitting defect $\Sigma$ vanishes if and only if some natural embedding keeps the orbitals of every site together, in particular whenever each site hosts one orbital.

\emph{Theorem.}  (i) For an isolated band with $C\neq0$,
\begin{equation}
\begin{aligned}
\det G_\star&=\pi^2C^2+\tfrac14\Torb^2,\\
\det(G_\star+\Sigma)&=\pi^2C^2+\tfrac14(\TLVPcl)^2 .
\end{aligned}
\label{eq:geom}
\end{equation}
(ii) Assume in addition that the band is exactly flat, that it is paired with its time-reversed partner by the on-site, orbital-diagonal attraction $-|U|\sum_{\bm R\alpha}n_{\bm R\alpha\uparrow}n_{\bm R\alpha\downarrow}$ (an attraction on the total density of a site with several orbitals adds inter-orbital terms, which part (ii) does not cover), and that pairing is uniform, meaning that the band has equal Brillouin-zone-averaged weight $n_\phi=(2\pi)^{-2}\int_\BZ|u_\alpha|^2\,\dd^2k$ on every orbital $\alpha$ that carries it.  Up to the term $-\frac12|U|n_\phi\hat N$, constant at fixed particle number $\hat N$, the band-projected Hamiltonian then has the $\eta$-pairing states, generated from the empty band by the zero-momentum pair creation operator $\eta^+$ of the band, as zero-energy ground states \cite{Yang1989,Tovmasyan2016,HHBK2024}.  If these are nondegenerate at fixed particle number, the exact zero-temperature stiffness in the thermodynamic limit is $\Ds(0)=\kappa_0G_\star$ \cite{HKH2026,WLY2026,SM}, the mean-field value \cite{PT2015,Huhtinen2022}, with $\kappa_0=(2/\pi^2)\Delta\sqrt{\nu(1-\nu)}$ at filling $\nu$ and pairing amplitude $\Delta=|U|n_\phi\sqrt{\nu(1-\nu)}$.  Relative to the topological floor $\det\Ds^{\rm floor}(0)=\kappa_0^2\pi^2C^2$ \cite{PT2015,OzawaMera2021,Prijon2026},
\begin{equation}
\frac{\det\Ds(0)}{\det\Ds^{\rm floor}(0)}=1+\Bigl(\frac{\Torb}{2\pi|C|}\Bigr)^2\le1+\Bigl(\frac{\TLVPcl}{2\pi|C|}\Bigr)^2,
\label{eq:central}
\end{equation}
with equality if and only if $\Sigma=0$.

Part (i) follows from Eqs.~\eqref{eq:detQ}--\eqref{eq:sigma}, the embedding-orbit lemma (End Matter), and the Loewner order of positive matrices; part (ii) follows from part (i) and $\Ds(0)=\kappa_0G_\star$.  The models of part (ii) are the simplest members of the quantum-geometric-nesting family \cite{HHBK2024,HKH2026}, for which two independent proofs show that the stiffness takes its BCS value when the paired ground state is nondegenerate \cite{HKH2026,WLY2026}; for a $C=\pm1$ band a rigorous bootstrap lower bound matches this value numerically \cite{GHK2026}.  The Supplemental Material (SM) \cite{SM} gives a short $\eta$-SU(2) argument for the exactness and exact diagonalization of the three models used below, which finds a nondegenerate ground state and the exact quadratic filling law of the stiffness in every sector computed.  The floor scales with the weight per active orbital, so that at fixed $|U|$ a band spread over more orbitals has a lower floor, which is why bands are compared at fixed pairing amplitude.  Exact flatness with $C\neq0$ needs hopping of unbounded range \cite{Chen2014}; the hypotheses hold for spectrally flattened bands, and approximately when the bandwidth is small compared with $|U|$, itself small compared with the gap.  Changes of the lattice shape at fixed cell area act as $G\to\mathsf MG\mathsf M^T$ with $\det\mathsf M=1$ \cite{WRS2026}, and $\min_{\mathsf M}\tr(\mathsf MG\mathsf M^T)=2\sqrt{\det G}$, so $(4\pi^2C^2+\Torb^2)^{1/2}$ is the least integrated quantum metric over all real-space realizations of the band, orbital positions and lattice shape included, and the floor is reached exactly when one of them is ideal.  All numerical values below were computed at two or more resolutions, and only agreeing digits are quoted \cite{SM}.

\paragraph{Consequences.}
First, a vortexable band has $\TLVPcl=\Torb=0$ and sits on the floor, which answers LVP's question for the stiffness.  Second, with $R\equiv\det\Ds/\det\Ds^{\rm floor}$, any enhancement costs vortexability,
\begin{equation}
\TLVPcl\ge2\pi|C|\sqrt{R-1},
\label{eq:tradeoff}
\end{equation}
which Fig.~\ref{fig:tradeoff} shows as a forbidden region.  Third, the converse, that a band on the floor is vortexable, can fail only through $\Sigma\neq0$, which requires internal components at one physical site (Cooper pairs relax them independently, a vortex coordinate cannot), or when $\TLVPcl=0$ is reached only at a collision of sites, because vortexability needs an actual coordinate.  If $\Torb=0$ while $\Sigma\neq0$, the vortex function $\hat z=\sum_{\bm R\alpha}\bm c\cdot(\bm R+\bm\delta_\alpha+\bm\eta^\star_\alpha)|\bm R\alpha\rangle\langle\bm R\alpha|$, with $\bm c$ the constant null vector of $\bar Q$ at the natural embedding and $\bm\delta_\alpha$ the reference positions, maps the band into itself but takes different values on the orbitals of one site, so its Jastrow factors do not screen contact repulsion between them.  LVP proposed such a layer-dependent operator for strained chiral twisted bilayer graphene \cite{LVP2023}.  In that model their operator does not map the band into itself, because the interlayer tunneling does not commute with it, and no layer-diagonal vortex function could map an exactly flat band with these layer spinors into itself (End Matter); numerically $\Torb>0$ there.  Fourth, in mean-field theory the natural embedding is encoded in the paired state.  With the relative phases $\theta_\alpha$ of the orbital order parameters relaxed at phase twist $\bm q$, the stiffness is the Schur complement $A_e-B_eH_e^{-1}B_e^T$ of the saddle Hessian, with $A_e=\kappa_0G_e$, so that $B_eH_e^{-1}B_e^T=\kappa_0\Gamma(\bm\eta-\bm\eta^\star)$, and $\dd\theta_\alpha/\dd\bm q=2(\bm\eta^\star_\alpha-\bm\eta_\alpha)$ up to a common constant (End Matter).

Two literature anchors place real models in Fig.~\ref{fig:tradeoff}.  The $\pi$-flux checkerboard band, studied by quantum Monte Carlo and the bootstrap \cite{HBC2020,GHK2026}, has its natural embedding at the physical positions and $R=1.163$; its stiffness exceeds the floor by $7.8\%$ along each principal axis.  Its orbitals sit on different sites, so $\Sigma=0$ and $\TLVP/2\pi=0.403$. It lies on the boundary curve, and its minimal metric reproduces the extrapolated bootstrap stiffness of Ref.~\cite{GHK2026} to a relative $7\times10^{-5}$.  The floor itself is reached by the honeycomb band of Li and Wu, a lattice model whose relative hoppings match a Wannier model of the two top moir\'e bands of twisted MoTe$_2$ \cite{LiWu2026}.  It samples the lowest Landau level at the two sublattice sites, is ideal with its orbitals on distinct sites and, by inversion, has equal sublattice weights; paired with its time-reversed partner it has $\Sigma=0$ and $R=1$, in mean-field theory and exactly if the paired ground state is nondegenerate \cite{SM}.  We do not suggest that the superconductivity of Ref.~\cite{Xu2025} is of this type.

\begin{figure}[t]
\includegraphics[width=\columnwidth]{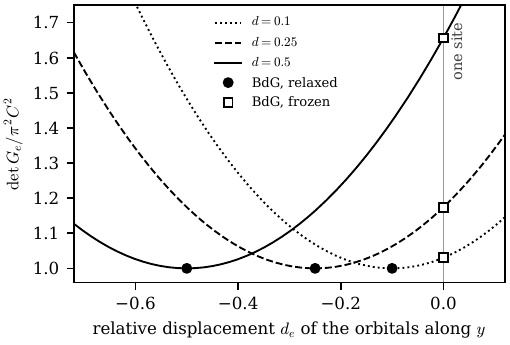}
\caption{Theta band, one set of Bloch states in two physical realizations.  Lines show $\det G_e/\pi^2C^2$ against the relative displacement $d_e$ of the two orbitals, for sampling points $d$ apart.  The minimum, $R=1$, is at $d_e=-d$, the physical embedding of the vortexable two-site realization; $d_e=0$ (vertical line) is the one-site realization.  Symbols show self-consistent BdG results ($96^2$ momenta, remote band at $80|U|$, half filling), which are the same for both realizations.  Filled circles give the relaxed stiffness, plotted at half the relative-phase response, and open squares the stiffness with phases frozen in the one-site frame.}
\label{fig:theta}
\end{figure}

\paragraph{The floor without vortexability.}
Ideal $|C|=1$ bands can be built by sampling the lowest Landau level on a lattice \cite{KM2010,DM2020,Wang2021}, also at two or more points per unit cell with ordinary translation symmetry \cite{LiWu2026,Zhao2026,WRS2026}.  We take the square-lattice instance with one flux quantum per cell and two sampling points a distance $d$ apart, and realize the same Bloch states in two ways (Fig.~\ref{fig:theta}).  With the orbitals on two sites at the sampling points, the physical embedding is natural and the band is vortexable, with $G_\star=\pi I$ and $R=1$.  With both orbitals on one site the stiffness is unchanged, because it does not depend on the embedding \cite{Huhtinen2022,TP2024}, but no common positions make the band vortexable.  Then $\TLVPcl/2\pi=0.174$, $0.417$, and $0.810$ for $d=0.1$, $0.25$, and $0.5$, and $\Sigma=w_{12}\bm d_\star\bm d_\star^T\neq0$ with $\bm d_\star=(0,-d)$ and $w_{12}=\int_\BZ w_1w_2\,\dd^2k$, so the bound \eqref{eq:central} is strict at $R=1$.  Self-consistent Bogoliubov--de Gennes (BdG) calculations for the flattened band confirm this, and the relative-phase response measures the sampling distance (Fig.~\ref{fig:theta}).  The exact Cooper-pair mass \cite{TLP2018,Iskin2021,Iskin2022,HA2026} and exact diagonalization, which finds the paired ground state nondegenerate on $3\times3$ and $3\times4$ tori, place the many-body stiffness on the floor as well, provided that nondegeneracy persists in the thermodynamic limit \cite{SM}, although the Berry curvature of this band fluctuates by $105\%$ (rms), as it may in ideal bands \cite{Estienne2023}. In the time-reversed channel ideality, not uniformity, decides.

Heterostrain gives a physical splitting defect.  We computed LVP's relaxed chiral model of twisted bilayer graphene, with equal and opposite relaxation strain in the two layers calibrated to the reported $80$~T peak pseudomagnetic field \cite{LVP2023,NK2017}.  Its flat band has $\TLVP/2\pi=2.540\times10^{-3}$, attained by a smooth common coordinate.  Layer-resolved displacements lower $\Torb/2\pi$ below $1.72\times10^{-4}$ (a variational bound) and, in a real-space solution, to about $2.3$--$2.4\times10^{-5}$, so $\Sigma\neq0$ (SM, Fig.~S2).  As LVP note, the strain gauge field alone leaves the band ideal, and the obstruction comes from the strained Fermi velocities, which differ between the layers \cite{LVP2023}; the End Matter adds that the gauge field alone keeps the layer spinors of all zero modes parallel.  The band density of twisted bilayer graphene is far from uniform, so this example concerns part (i) of the theorem, not the stiffness.

\paragraph{Continuum.}
For a scalar continuum band a periodic displacement field acts as $u_{\bm k}(\bm r)\to\e^{-\ii\bm k\cdot\bm\eta(\bm r)}u_{\bm k}(\bm r)$.  The Gram structure survives on the Hilbert completion of the covariance form, $G[\bm\eta]\succeq G^{\inf}$ with $\inf_{\bm\eta}\det G[\bm\eta]=\det G^{\inf}$ (End Matter).  We prove that lifted displacements of torus diffeomorphisms isotopic to the identity are $L^p$ dense among periodic displacement fields \cite{SM} (compare Ref.~\cite{Brenier2003} for bounded domains), so the relaxed LVP infimum equals $\det G^{\inf}$, and a scalar band has no splitting defect.  For an isolated flat time-reversed band with contact attraction and uniform pairing, which here requires a uniform integrated density $\bar w(\bm r)>0$, the covariance operator is Fredholm, the minimizer $\bm\eta^\star$ is smooth, and the mean-field stiffness is $\kappa_0G^{\inf}$ \cite{ZM2025,SM}.  Hence, in mean-field theory,
\begin{equation}
\frac{\det\Ds(0)}{\det\Ds^{\rm floor}(0)}=1+\Bigl(\frac{\TLVPcl}{2\pi|C|}\Bigr)^2 ,
\label{eq:continuum}
\end{equation}
and, when the covariance form has only translations as zero modes, the infimum is attained by an LVP coordinate if and only if $\det(I+\nabla\bm\eta^\star)>0$ everywhere (End Matter).  Landau levels lie on the boundary curve. The $n$th level has $G=\pi(2n+1)I$ \cite{PT2015}, its native frame is natural, so that $G_\star=G$ \cite{SM}, and $\TLVP=\Torb=4\pi\sqrt{n(n+1)}$, so $R=(2n+1)^2$ in mean field (for $n\ge1$ compare the higher vortexability of Ref.~\cite{Fujimoto2025}).

\begin{figure}[t]
\includegraphics[width=\columnwidth]{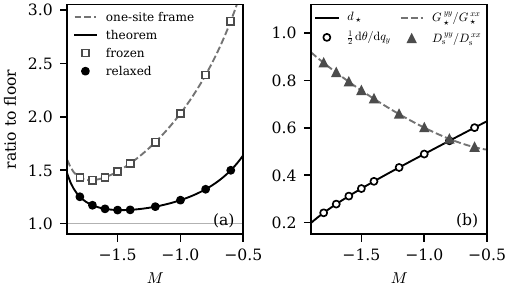}
\caption{Test on the QWZ band, Eq.~\eqref{eq:qwzx}.  (a) $R$ from the theorem (solid) and $\det G/\pi^2C^2$ in the one-site frame, which equals $1+(\TLVPcl/2\pi|C|)^2$ of the one-site realization (dashed); self-consistent BdG ($96^2$ momenta, remote band at $80|U|$, half filling) with relaxed (filled circles) or frozen (open squares) order parameters.  (b) Natural displacement $d_\star$ (solid) and half the relative-phase response (circles); anisotropy of $G_\star$ (dashed) and of the BdG stiffness (triangles).}
\label{fig:qwz}
\end{figure}

\paragraph{Test away from saturation.}
To test Eq.~\eqref{eq:central} at $R>1$ we use the Qi--Wu--Zhang (QWZ) band \cite{QWZ2006} in the orbital basis that makes the mass term off-diagonal,
\begin{equation}
h(\bm k)=(M-\cos k_x-\cos k_y)\sigma_x-\sin k_y\,\sigma_y+\sin k_x\,\sigma_z ,
\label{eq:qwzx}
\end{equation}
whose lower band has $|C|=1$ for $-2<M<0$.  Since $\sigma_xh(\bm k)\sigma_x=h(-\bm k)$, the integrated orbital weights are equal and uniform pairing holds for every $M$.  The natural embedding separates the orbitals by $\bm d_\star=(0,d_\star)$; at $M=-1$, $d_\star=0.489$ lattice constants and $\det G_\star/\pi^2=1.2195$, against $2.031$ in the one-site frame.  Self-consistent BdG calculations for the spectrally flattened band, with the remote band at $E_g=80|U|$, follow the theorem across the family to $3\times10^{-5}$, a residual that scales as $(|U|/E_g)^2$, and not the one-site value (Fig.~\ref{fig:qwz}).  The phase response measures $d_\star$, and the stiffness is anisotropic, $\Ds^{yy}/\Ds^{xx}=0.60$ at $M=-1$, although the band is $C_4$ symmetric, because in this basis $C_4$ mixes the orbitals and the orbital-diagonal attraction breaks it.  In the standard basis, where $C_4$ acts by orbital phases, the natural embedding is on site, as the symmetry theorem of Ref.~\cite{WRS2026} requires, but uniform pairing fails \cite{SM}.

\paragraph{Finite temperature.}
Under the hypotheses of part (ii) the projected model has an exact $\eta$-pairing SU(2) symmetry \cite{Tovmasyan2016}.  Its paired state is a pseudospin ferromagnet with a quadratic Goldstone mode \cite{HHBK2024,HKH2026}, and in two dimensions there is no finite-temperature transition \cite{MerminWagner1966}, although $\Ds(0)>0$.  A finite $T_c$ needs a symmetry-breaking anisotropy, of energy scale $\varepsilon_a$, and its sign matters. In quantum Monte Carlo on the $\pi$-flux band, interband processes and residual bandwidth favor superconductivity, whereas a weak nearest-neighbor attraction drives phase separation \cite{HBC2020}.  For a weak easy-plane anisotropy, Ref.~\cite{HHBK2024} estimates $k_BT_c\simeq4\pi|\Upsilon|/\ln(|\Upsilon|/\varepsilon_a)$, with $|\Upsilon|=\sqrt{\det\Upsilon}$ the geometric mean of the pseudospin stiffness $\Upsilon$, from the O(3)-to-XY crossover \cite{NP1977,HikamiTsuneto1980}.  Since $\Upsilon$ is proportional to $\Ds(0)$ at fixed filling, bands compared at equal pairing amplitude, filling, and $\varepsilon_a$ obey $(T_c/T_c^{\rm floor})^2=R\,[1+O(1/\ln(|\Upsilon|/\varepsilon_a))]$, and exactly $R$ at this order if $\varepsilon_a$ scales with $|\Upsilon|$ (End Matter).  We make no claim for strong symmetry breaking.

\paragraph{Outlook.}
The identity extends in weaker forms \cite{SM}.  With dispersion the mean-field stiffness becomes a weighted Gram problem, in which two-sided bounds in $\Torb$ survive, but vortexability no longer pins the stiffness.  Without uniform pairing the relevant geometry is that of the pairing-deformed band, which keeps the Chern number and hence a floor; this is a mean-field statement, since such models are not quantum-geometric nesting models \cite{HHBK2024}.  With spin-orbit coupling, time reversal makes the spin locking of singlet phases free.  Self-consistent BdG calculations confirm all three.  Pairing within a single Chern band behaves differently. A uniform Landau level has no same-band stiffness \cite{ZM2025,MC2024,GAF2025}, inhomogeneous ideal bands can have one \cite{GAF2025}, bounded above in Ref.~\cite{MC2024}, and outside the time-reversed class the BCS value need not be exact \cite{HHGK2025,HKH2026}.  The ordering concerns on-site attraction. Other nesting interactions shift the stiffness of a given band \cite{GHK2026}, and for interlayer exciton condensates with gate-screened Coulomb interactions pure Landau levels are stiffer than mixtures with the same quantum geometry \cite{Zou2026}, whereas under component-diagonal contact attraction we predict that the pure $n=1$ level and the equal mixture of $n=0$ and $n=2$ have the same $R=9$ and, at equal pairing amplitude, the same mean-field stiffness.  As Zou \emph{et al.} note \cite{Zou2026}, a particle-hole transformation of one layer maps our model onto an interlayer exciton condensate in a bilayer whose layers carry the same flat Chern band, up to constant orbital phases, at total filling one, with on-site interlayer interaction \cite{VGQ2024}, and the counterflow vector potential onto a common one.  Under the same hypotheses (uniform orbital weights, nondegenerate ground state), Eq.~\eqref{eq:central} and the floor therefore also hold for the exact counterflow stiffness of such bilayers, zero-field analogs of quantum Hall bilayers.  Finally, a small $\TLVP$ is evidence for, not proof of, the stability of fractional Chern insulators \cite{Jackson2015,Moitra2026}; our results concern band geometry and the paired state, not the energetics of competing phases.

In summary, for time-reversed flat-band superconductors with uniform pairing and a nondegenerate paired ground state, the minimal quantum metric that fixes the exact stiffness and the frame-optimized metric that measures LVP vortexability are minima of one quadratic form over nested spaces.  Vortexable bands therefore sit on the topological floor of the stiffness, the cost of exceeding it is paid in vortexability, and the converse fails through the splitting defect of internal components that share a site, which the theta band realizes exactly and heterostrained twisted bilayer graphene realizes, in its band geometry, in a physical band.

\begin{acknowledgments}
Claude Opus 5.5 (Anthropic) was used to assist with calculations and numerical simulations. The authors verified all results and take full responsibility for the content of this work.
\end{acknowledgments}

\textit{Data availability.}  The data that support the findings of this article, together with the scripts that produce every figure, table, and number in this Letter and its Supplemental Material, are openly available \cite{Zenodo}.

\clearpage
\section*{End Matter}

\paragraph{Embedding-orbit lemma.}
Let $\mathbb T^2=\mathbb R^2/\Lambda$, and let $p$ and $p'$ be ordered collision-free configurations of the $N_{\rm s}$ site positions.  The ordered configuration space of $\mathbb T^2$ is path connected, so choose a smooth collision-free path $p(t)$ from $p$ to $p'$.  Regarding the labeled points as a compact zero-dimensional submanifold, the isotopy-extension theorem extends $p(t)$ to an ambient isotopy $F_t$ of $\mathbb T^2$ with $F_0=\mathrm{id}$ \cite{Palais1960}.  Because $F_t$ is isotopic to the identity, it acts trivially on $\pi_1(\mathbb T^2)=\Lambda$, and its lift with $\widetilde F_0=\mathrm{id}$ obeys $\widetilde F_t(\bm r+\bm R)=\widetilde F_t(\bm r)+\bm R$; it is orientation preserving.  Independent lattice shifts of the labeled representatives are produced by letting a path wind around the fundamental cycles while avoiding the other points.  Every collision-free site-rigid embedding is therefore reached by a lattice-periodic diffeomorphism, and conversely by evaluation at the sites.  Collision-free configurations are dense in $\Vcal$ and $G_e$ is quadratic in the displacements, so $\inf_\phi\det\bar Q_\phi=\min_{e\in\Vcal}\det\bar Q_e$.

\paragraph{Gram identity.}
Let $\hat\eta_j=\mathrm{diag}(\eta_{\alpha j})$.  Since $u_e=\mathcal U_eu$ with $\mathcal U_e=\mathrm{diag}(\e^{-\ii\bm k\cdot\bm\eta_\alpha})$, one has $(1-P_e)\partial_ju_e=\mathcal U_e(1-P)(\partial_j-\ii\hat\eta_j)u$, and the real part of $\langle\partial_iu_e|(1-P_e)|\partial_ju_e\rangle$ is
\begin{multline}
g^e_{ij}=g_{ij}+\sum_\alpha(\eta_{\alpha i}\beta_{\alpha j}+\eta_{\alpha j}\beta_{\alpha i})\\
+\sum_{\alpha\beta}\eta_{\alpha i}(w_\alpha\delta_{\alpha\beta}-w_\alpha w_\beta)\eta_{\beta j},
\label{eq:shift}
\end{multline}
with $\beta_{\alpha j}$ the integrand of $b_{\alpha j}$ \cite{TP2024}.  The imaginary part changes by a total derivative, so $C$ is unchanged.  Let $\bm z\in\ker W$ and set $\bm\eta_j=t\bm z$ in one direction $j$ only. Then $G_{jj}(t)=G_{jj}(0)+2t\,\bm z^T\bm b_j$, which is non-negative for all real $t$ only if $\bm z^T\bm b_j=0$.  Hence $\bm b_j\perp\ker W$, that is, $\bm b_j\in\operatorname{Ran}W$, the normal equations $W\bm\eta^\star_j=-\bm b_j$ are solvable even when $W$ has zero modes other than translations, and completing the square gives Eq.~\eqref{eq:gram}.  For real $\bm v$, $\bm v^T\Gamma(\bm\zeta)\bm v$ is the Brillouin-zone integral of the variance of $\sum_jv_j\zeta_{\alpha j}$ in the distribution $w_\alpha$.  Since $C\neq0$ forces $G_\star\succ0$, the Loewner order implies $\det G_e\ge\det G_\star$, with equality only for $\Gamma=0$.  Writing $u_\alpha=\sqrt{w_\alpha}\,\e^{\ii\Phi_\alpha}$, the metric is a quarter of the Fisher metric of $w_\alpha$, which no embedding changes, plus the covariance of the phase gradients $\partial_j\Phi_\alpha$, which an embedding shifts by $-\eta_{\alpha j}$. The natural embedding is therefore the least-squares fit of constant positions to these gradients.

\paragraph{Free-energy Hessian.}
Let $\mathcal F_e(\bm q,\xi)$ be the mean-field grand potential at phase twist $\bm q$ in the embedding $e$, with $\xi$ the internal variables (amplitudes and relative phases $\theta_\alpha$ of the orbital order parameters, chemical potential), and $A_e$, $B_e$, $H_e$ its second derivatives in $\bm q\bm q$, $\bm q\xi$, and $\xi\xi$ at the saddle.  Differentiating $\partial_\xi\mathcal F_e(\bm q,\xi(\bm q))=0$ gives $\dd\xi/\dd\bm q=-H_e^{-1}B_e^T$ and the relaxed Hessian $A_e-B_eH_e^{-1}B_e^T$.  Time reversal makes the amplitudes and the chemical potential even in $\bm q$, and the grand potential at $\bm q=0$ even in the relative phases, so only the relative phases enter $B_e$ and $H_e$ does not couple them to the other variables.  The twist enters only through the phase of the pair field, as in the BdG calculations of the SM.  For the pair field $\Delta_\alpha\e^{\ii\theta_\alpha}\e^{2\ii\bm q\cdot(\bm R+\bm\delta_\alpha+\bm\eta_\alpha)}$, shifting the phases by $\theta_\alpha\to\theta_\alpha+2\bm q\cdot\bm\zeta_\alpha$ is the same as moving the embedding by $\bm\zeta$, so $\mathcal F_e(\bm q,\theta+2\bm\zeta\bm q)=\mathcal F_{e+\zeta}(\bm q,\theta)$ with the other internal variables held fixed.  In the flat uniform-pairing limit the frozen Hessian is $A_{e'}=\kappa_0G_{e'}$ in every embedding $e'$ \cite{PT2015,SM}.  Expanding both sides to second order in $\bm q$ and inserting Eq.~\eqref{eq:gram} gives $B_e=\frac12\kappa_0\bm b_e^T$ and, on the relative phases, $H_e=\frac14\kappa_0W$, where $\bm b_e=\bm b+W\bm\eta$ is the linear coefficient of $G_e$.  At the natural embedding $e_\star$ ($\bm\eta=\bm\eta^\star$), $\bm b_{e_\star}=0$, so $B_{e_\star}=0$; in general, with $H_e^{-1}$ read as the pseudoinverse on the relative phases, $A_e-B_eH_e^{-1}B_e^T=\kappa_0(G_e-\bm b_e^TW^+\bm b_e)=\kappa_0G_\star$ and $\dd\theta/\dd\bm q=-2W^+\bm b_e=2(\bm\eta^\star-\bm\eta)$ up to a common constant.

\paragraph{Heterostrain.}
In LVP's chiral model \cite{LVP2023} the layer blocks of the zero-mode operator $D$ are $2\chi^\mu_\pm(-\ii\partial_\mu+A_{\pm\mu})$, coupled by the tunneling $\alpha_{\rm w}\mathcal T(\pm\bm r)$; the tetrad $\chi_\pm$ and the gauge field $\bm A_\pm$ come from the relaxation strain, equal and opposite in the two layers.  \emph{Proposition.}  Without the tetrad strain the layer blocks are $-2\ii\bar\partial+a_\pm(\bm r)$, with $\bar\partial=\frac12(\partial_x+\ii\partial_y)$ and $a_\pm=A_{\pm x}+\ii A_{\pm y}$.  (a) $[D,z\mathbf 1]=0$ for $z=x+\ii y$, so the zero-mode band, where it is an exactly flat band, is vortexable \cite{LVP2023}; (b) if $a_++a_-=0$, the layer spinors of all zero modes are parallel at every point.  \emph{Proof.}  (a) $[-2\ii\bar\partial+a_\pm,z]=-2\ii\bar\partial z=0$, and the tunneling commutes with $z\mathbf 1$.  (b) For zero modes $\psi,\psi'$ at momenta $\bm k,\bm k'$, the Wronskian $\mathcal W=\psi_+\psi'_--\psi_-\psi'_+$ obeys $\bar\partial\mathcal W=(a_++a_-)\mathcal W/2\ii=0$, since the tunneling terms cancel in pairs.  It is quasiperiodic and bounded, hence constant, and zero unless $\bm k+\bm k'+\bm q_1$ is a reciprocal vector ($\bm q_1$ the momentum transfer of the tunneling, SM Sec.~S6.1), a case that continuity in $\bm k'$ covers.  $\square$  With the tetrad the layer spinors span both directions (numerically, the layer matrix $\mathcal L(\bm r)=\int_\BZ\psi_{\bm k}(\bm r)\psi_{\bm k}(\bm r)^\dagger\dd^2k$ has rank two, with eigenvalue ratio up to $1.8\times10^{-5}$; zero modes of $D$ are real analytic, so $\det\mathcal L$ vanishes only on a set of measure zero).  A vortex function diagonal in the layer index that maps an exactly flat band into itself would then have to be common to both layers and annihilated by both $\chi_+^\mu\partial_\mu$ and $\chi_-^\mu\partial_\mu$; at first order in strain these are independent wherever the traceless strain is nonzero, as LVP note for a common vortex function \cite{LVP2023}.  For the first-harmonic relaxation field the traceless strain vanishes only on a set of measure zero, so the gradient of such a function would vanish almost everywhere, and it could not grow across the lattice as a vortex function must.  This makes $\Torb>0$ the expected outcome; the computation \cite{SM} finds it, for a band whose residual bandwidth is $1.1\times10^{-6}\,vk_\theta$, with $v$ the Dirac velocity and $k_\theta$ the moir\'e wave number.

\paragraph{Leading-log corollary.}
With $|\Upsilon|=\sqrt{\det\Upsilon}\propto\sqrt{\det\Ds(0)}$, $T_c/T_c^{\rm floor}=\sqrt R\,\ln(|\Upsilon_{\rm floor}|/\varepsilon_a)/\ln(\sqrt R|\Upsilon_{\rm floor}|/\varepsilon_a)=\sqrt R\,[1-\ln\sqrt R/\ln(|\Upsilon_{\rm floor}|/\varepsilon_a)+\dots]$ at fixed $\varepsilon_a$, which gives the stated ratio; if $\varepsilon_a\propto|\Upsilon|$ the logarithms are equal and $T_c/T_c^{\rm floor}=\sqrt R$.  For $\varepsilon_a\to0$ thermal depletion of the stiffness drops out because $T_c\to0$.

\paragraph{Continuum.}
For fields $f,g$ on the unit cell let
$$\mathfrak w(f,g)=\int_\BZ\Bigl[\int w_{\bm k}fg-\int w_{\bm k}f\int w_{\bm k}g\Bigr]\dd^2k,$$
with $w_{\bm k}=|u_{\bm k}(\bm r)|^2$, be the covariance form, $\ell_j$ the linear term, and $G^0$ the metric of the undisplaced band, so that $G_{ij}[\bm\eta]=G^0_{ij}+\ell_j(\eta_i)+\ell_i(\eta_j)+\mathfrak w(\eta_i,\eta_j)$.  Positivity of $G_{jj}[tf]$ for all real $t$ gives $|\ell_j(f)|^2\le G^0_{jj}\,\mathfrak w(f,f)$, so $\ell_j$ is continuous in the covariance seminorm.  On the Hilbert completion of the fields modulo the null space of $\mathfrak w$, with canonical map $J$, the Riesz representation $\ell_j(f)=\langle\mathfrak h_j,Jf\rangle$ gives $G_{ij}[\bm\eta]=G^{\inf}_{ij}+\langle J\eta_i+\mathfrak h_i,J\eta_j+\mathfrak h_j\rangle$ with $G^{\inf}_{ij}=G^0_{ij}-\langle\mathfrak h_i,\mathfrak h_j\rangle$.  With uniform density $\bar w$ the covariance operator is $W=\bar w(1-\widetilde K)$, where $\widetilde K$ has the kernel $\bar w^{-1}\int_\BZ w_{\bm k}(\bm r)w_{\bm k}(\bm r')\,\dd^2k$ and is compact and smoothing, and the Fredholm alternative gives a smooth minimizer, unique modulo $\ker W$.  The minimizer defines an LVP coordinate when $F^\star=\mathrm{id}+\bm\eta^\star$ is an orientation-preserving diffeomorphism.  If $\ker W$ contains only translations, this holds if and only if $\det(I+\nabla\bm\eta^\star)>0$, since $F^\star$ is then a local diffeomorphism of degree one, hence global, and compactness gives LVP's uniform nondegeneracy.  The stiffness relation uses the contact-interaction result of Ref.~\cite{ZM2025} for every twist direction and polarization of the resulting quadratic form \cite{SM}; it is a mean-field statement.

\end{document}


\title{Supplemental Material for ``Vortexable Chern Bands Sit Exactly on the Topological Floor of Flat-Band Superfluid Stiffness''}

\author{Zhi-Wei Wang}
\email{zhiweiwang.phy@gmail.com}
\affiliation{College of Physics, Jilin University, Changchun, 130012, People's Republic of China}
\affiliation{Department of Computer Science, University of York, York YO10 5GH, United Kingdom}

\author{Samuel L. Braunstein}
\email{sam.braunstein@york.ac.uk}
\affiliation{Department of Computer Science, University of York, York YO10 5GH, United Kingdom}

\maketitle

\setcounter{equation}{0}
\renewcommand{\theequation}{S\arabic{equation}}
\setcounter{figure}{0}
\renewcommand{\thefigure}{S\arabic{figure}}
\setcounter{table}{0}
\renewcommand{\thetable}{S\arabic{table}}

This Supplemental Material fixes conventions and gives the derivations and numerical evidence behind the Letter.  Section~S1 sets conventions.  Sections~S2 and S3 prove the geometric statements: the relation between LVP coordinate frames and orbital embeddings, the Gram identity, and the splitting defect.  Section~S4 treats the stiffness: its normalization, the exactness argument, exact diagonalization, the exact Cooper-pair mass, and the free-energy Hessian.  Section~S5 collects the examples (theta, $\pi$-flux, honeycomb, elliptic, Landau-level, and QWZ bands) with their tests, and Sec.~S6 the computation for heterostrained chiral twisted bilayer graphene.  Section~S7 proves the continuum statements, Sec.~S8 discusses finite temperature, Sec.~S9 the robustness of the identity, and Sec.~S10 the exciton dual.  Section~S11 records a pointwise decomposition of the integrated metric, and Sec.~S12 lists the scripts.  Statements labeled ``numerical'' were computed at two or more resolutions, and only agreeing digits are quoted.  Equation and figure numbers without the prefix S refer to the Letter.

\section*{S1. Conventions}

For a normalized isolated-band Bloch vector $u(\bm k)$ with projector $P(\bm k)=|u\rangle\langle u|$ we use
\begin{equation}
Q_{ij}(\bm k)=\Tr\bigl[P\,\partial_iP\,\partial_jP\bigr]=g_{ij}(\bm k)-\frac{\ii}{2}\Om_{ij}(\bm k),\qquad
C=\frac1{2\pi}\int_\BZ\Om_{xy}\,\dd^2k,\qquad
G=\int_\BZ g\,\dd^2k,
\label{S:qgt}
\end{equation}
with $\partial_i=\partial/\partial k_i$ and the measure $\dd^2k$ without a factor $(2\pi)^{-2}$.  Then $\bar Q=\int_\BZ Q\,\dd^2k=G-\ii\pi C\mathsf J$ with $\mathsf J_{xy}=-\mathsf J_{yx}=1$, so that
\begin{equation}
\det\bar Q=\det G-\pi^2C^2 .
\label{S:detQ}
\end{equation}
The determinant does not depend on the transpose convention of Ref.~\cite{LVP2023}, and $\det G/\pi^2C^2$ does not change under linear reparametrizations of $\bm k$, so it can be evaluated in lattice coordinates.  Positivity of $Q$ gives $2\sqrt{\det g}\ge|\Om_{xy}|$ pointwise, and Minkowski's determinant inequality for positive $2\times2$ matrices gives
\begin{equation}
\sqrt{\det G}\ge\int_\BZ\sqrt{\det g}\,\dd^2k\ge\frac12\int_\BZ|\Om_{xy}|\,\dd^2k\ge\pi|C|,
\label{S:chain}
\end{equation}
the determinant form of the Chern bound \cite{PT2015,OzawaMera2021,Prijon2026}.  Equality holds in the first step if and only if $g(\bm k)$ is proportional to one constant matrix almost everywhere (the equality case of Minkowski's inequality), in the second if and only if $2\sqrt{\det g}=|\Om_{xy}|$ almost everywhere, and in the third if and only if $\Om_{xy}$ has one sign.  Equality throughout therefore holds if and only if $g=\frac12|\Om_{xy}|\,\omega$ almost everywhere with a constant unimodular $\omega>0$ and $\Om_{xy}$ of one sign.

For two-band models with $P=(1-\bm n\cdot\bm\sigma)/2$ one has $g_{ij}=\frac14\partial_i\bm n\cdot\partial_j\bm n$ and $\Om_{xy}=\frac12\bm n\cdot(\partial_x\bm n\times\partial_y\bm n)$, hence $2\sqrt{\det g}=|\Om_{xy}|$ pointwise.

\paragraph{Stiffness normalization.}  The zero-temperature superfluid stiffness tensor $\Ds$ is the second derivative of the ground-state energy per unit area with respect to a uniform vector potential, which twists every electron as $\bm k\to\bm k+\bm A$ in units with $\hbar$ and the electron charge set to one.  For the flat-band models of the Letter, with unit cell area one,
\begin{equation}
\Ds(0)=\kappa_0G_\star,\qquad
\kappa_0=\frac{2}{\pi^2}\,\Delta\sqrt{\nu(1-\nu)}=\frac{2}{\pi^2}\,|U|n_\phi\,\nu(1-\nu),\qquad
\Delta=|U|n_\phi\sqrt{\nu(1-\nu)},
\label{S:kappa0}
\end{equation}
where $\nu$ is the filling of the flat band, $G_\star$ the natural (Loewner-minimal) integrated metric of Eq.~\eqref{S:gram}, $n_\phi$ the Brillouin-zone-averaged weight of the band per active orbital (Sec.~S4.1), and $\Delta$ the pairing amplitude.  Equivalently $\Ds=8|U|n_\phi\,\nu(1-\nu)\,\gbar$ with $\gbar=G_\star/4\pi^2$ the Brillouin-zone average of the natural metric.  Other normalizations in the literature differ by constant factors, for example $\Ds=E''/(4N_{\rm c})$ in Ref.~\cite{GHK2026}, with $E''$ the second derivative of the ground-state energy with respect to the vector potential and $N_{\rm c}$ the number of unit cells, which gives $|U|\gbar\,\nu(1-\nu)$ for $n_\phi=1/2$; all of them cancel in the ratio $R=\det\Ds/\det\Ds^{\rm floor}$, with $\det\Ds^{\rm floor}=\kappa_0^2\pi^2C^2$.  In the numerical BdG work Hessians are quoted in units of $\kappa_H=2|U|n_\phi\nu(1-\nu)/\pi^2=|U|/4\pi^2$ ($n_\phi=\nu=1/2$), so that $\det(\cdot)/(\kappa_H\pi|C|)^2$ is $R$ with no fitted parameter.

\section*{S2. LVP frames and point-orbital embeddings}

\subsection*{S2.1 The LVP measure and its closure}
LVP define a coordinate-dependent positive-semidefinite quantum geometric tensor ($\tilde\eta$ in their notation, $\widetilde Q_\phi(\bm k)$ here), its integral $\widetilde Q_{\phi,I}=\int_\BZ\widetilde Q_\phi\,\dd^2k$ [their Eq.~(38)], and the measure [their Eq.~(39)]
\begin{equation}
\TLVP=\min_{\phi}2\sqrt{\det\widetilde Q_{\phi,I}},
\label{S:TLVP}
\end{equation}
where $\phi$ runs over lattice-periodic orientation-preserving diffeomorphisms [their Eq.~(35)], $\phi(\bm r+\bm R)=\phi(\bm r)+\bm R$ for $\bm R\in\Lambda$ \cite{LVP2023}.  A band is vortexable with a vortex function of lattice-periodic gradient if and only if $\TLVP=0$.  Their index order transposes ours, so $\det\widetilde Q_{\phi,I}=\det\bar Q_\phi=\det G_\phi-\pi^2C^2$ in the frame $\phi$.  LVP note that this modification of band geometry had been interpreted, for tight-binding models, as a choice of unit-cell embedding \cite{SimonRudner2020,Huhtinen2022}.  Because the minimum need not be attained, we use the closure value
\begin{equation}
\TLVPcl=2\inf_{\phi\in\Dcal}\sqrt{\det\bar Q_\phi},
\label{S:TLVPcl}
\end{equation}
with $\Dcal$ the set of lattice-periodic orientation-preserving diffeomorphisms.  $\TLVPcl=\TLVP$ whenever the infimum is attained by an admissible coordinate.  Vortexability implies $\TLVPcl=0$; the converse needs attainment by a nondegenerate vortex coordinate, which LVP's argument for exact short-range ground states requires.

Away from zero, the determinant measure should not be identified with a fixed-frame trace defect.  If symmetry makes $G=aI$, the trace excess is $2a-2\pi|C|$ while the determinant measure is $2(a^2-\pi^2C^2)^{1/2}$; the two have the same zero but are different functions.

\subsection*{S2.2 Point orbitals see a frame only at their positions}
Let the $N$ orbitals of a unit cell sit at $\bm\delta_\alpha$.  LVP define their measure for continuum Bloch functions; for a tight-binding model we define it by treating the orbitals as points, which is a modeling choice.  In this strict point-orbital limit $\langle\bm r|\bm R,\alpha\rangle\propto\delta^{(2)}(\bm r-\bm R-\bm\delta_\alpha)$, and for $\phi\in\Dcal$, $\phi(\bm R+\bm\delta_\alpha)=\bm R+\phi(\bm\delta_\alpha)$.  The LVP periodic Bloch factor $\exp[-\ii\bm k\cdot\phi(\bm r)]$ therefore acts on orbital $\alpha$ as the re-embedding $\bm\delta_\alpha\mapsto\phi(\bm\delta_\alpha)$,
\begin{equation}
P_e(\bm k)=\mathcal U_e(\bm k)P_0(\bm k)\mathcal U_e(\bm k)^\dagger,\qquad \mathcal U_e(\bm k)=\mathrm{diag}\bigl(\e^{-\ii\bm k\cdot\bm\eta_1},\ldots,\e^{-\ii\bm k\cdot\bm\eta_N}\bigr),\qquad \bm\eta_\alpha=\phi(\bm\delta_\alpha)-\bm\delta_\alpha .
\label{S:Ue}
\end{equation}
Orbitals at the same position receive the same displacement, so every frame acts as a site-rigid embedding.

\subsection*{S2.3 Converse: configuration spaces and isotopy extension}
Let the labeled points now be the $N_{\rm s}$ sites.  Modulo $\Lambda$, a collision-free labeled embedding is a point of the ordered configuration space $F_{N_{\rm s}}(\mathbb T^2)$, which is path connected because $\mathbb T^2$ is a connected manifold of dimension two.  A smooth collision-free path $p_s(t)$ between two such embeddings is an isotopy of a compact zero-dimensional submanifold, and the isotopy-extension theorem extends it to an ambient isotopy $F_t$ of $\mathbb T^2$ with $F_0=\mathrm{id}$ and $F_t[p_s(0)]=p_s(t)$ \cite{Palais1960}.  Because $F_t$ is isotopic to the identity it acts trivially on $\pi_1(\mathbb T^2)=\Lambda$, so its lift with $\widetilde F_0=\mathrm{id}$ obeys $\widetilde F_t(\bm r+\bm R)=\widetilde F_t(\bm r)+\bm R$, and it is orientation preserving.  Independent lattice shifts of the labeled representatives are produced by letting a strand wind around either fundamental cycle of the punctured torus while avoiding the remaining finitely many points.  Hence
\begin{equation}
\{\text{LVP frames, evaluated at the sites}\}=\{\text{collision-free site-rigid embeddings}\}.
\label{S:orbit}
\end{equation}

\subsection*{S2.4 Collisions and the closure}
A diffeomorphism cannot merge two sites, but the algebraic site-rigid minimum may lie on a collision locus.  From Eq.~\eqref{S:Ue}, $G_e$ is a quadratic function of the displacements, and collision-free embeddings are dense in the site-rigid space $\Vcal$, so
\begin{equation}
\inf_{\phi\in\Dcal}\det G_\phi=\min_{e\in\Vcal}\det G_e .
\label{S:closure}
\end{equation}
This is the content of $\TLVPcl=2\min_{e\in\Vcal}\sqrt{\det\bar Q_e}$ in the Letter.  If the site-rigid minimizer is collision free, $\TLVPcl=\TLVP$.  Several orbitals sharing a site move together under every frame, so the LVP orbit lies in $\Vcal$; the paired state has no such constraint, because every orbital carries its own order parameter.

\section*{S3. Gram identity, Loewner minimality, and the splitting defect}

The integrated metric is quadratic in the orbital positions; this transformation law is in Ref.~\cite{TP2024} [their Eq.~(89)] and, with the minimality of the natural embedding in every direction (equivalently, in the Loewner order), its uniqueness up to translation, and its symmetry, in Ref.~\cite{WRS2026}; Ref.~\cite{Huhtinen2022} showed that orbitals at high-symmetry positions give the minimal metric.  We write the law as a Gram identity, the completed square behind that minimality, and restrict it to site-rigid embeddings; the restriction, and with it the splitting defect $\Sigma$, is new.

\subsection*{S3.1 Gram identity}
Fix a reference embedding with Bloch vector $u$ and let an embedding $e$ displace orbital $\alpha$ by $\bm\eta_\alpha$.  With $\hat\eta_j=\mathrm{diag}(\eta_{\alpha j})$,
\begin{equation}
u_e=\mathcal U_eu,\qquad (1-P_e)\partial_ju_e=\mathcal U_e(1-P)(\partial_j-\ii\hat\eta_j)u .
\end{equation}
With $v_j=(1-P)\partial_ju$ and $w_\alpha=|u_\alpha|^2$, the real part of $Q^e_{ij}=\langle(\partial_i-\ii\hat\eta_i)u|(1-P)|(\partial_j-\ii\hat\eta_j)u\rangle$ is
\begin{equation}
g^e_{ij}=g_{ij}+\sum_\alpha(\eta_{\alpha i}\beta_{\alpha j}+\eta_{\alpha j}\beta_{\alpha i})+\sum_{\alpha\beta}\eta_{\alpha i}(w_\alpha\delta_{\alpha\beta}-w_\alpha w_\beta)\eta_{\beta j},\qquad
\beta_{\alpha j}=-\mathrm{Im}\bigl[u_\alpha^*v_{j\alpha}\bigr].
\label{S:gshift}
\end{equation}
The quadratic term is $\langle\hat\eta_i\hat\eta_j\rangle-\langle\hat\eta_i\rangle\langle\hat\eta_j\rangle$, the covariance of the displacements in the distribution $w_\alpha(\bm k)$.  The Berry connection shifts by $\langle\hat\eta_j\rangle$, so the local curvature changes by a curl and $C$ does not change.  Integrating,
\begin{equation}
G_e=G+\bm\eta^T\bm b+\bm b^T\bm\eta+\bm\eta^TW\bm\eta,\qquad
W_{\alpha\beta}=\int_\BZ(w_\alpha\delta_{\alpha\beta}-w_\alpha w_\beta)\,\dd^2k,\qquad
b_{\alpha j}=\int_\BZ\beta_{\alpha j}\,\dd^2k,
\label{S:Gquad}
\end{equation}
with $\bm\eta$ and $\bm b$ arrays of size $N\times2$.  $W$ is real, symmetric, and positive semidefinite, and uniform translations lie in its kernel.  Other zero modes cause no difficulty: for $\bm z\in\ker W$ and $\bm\eta_j=t\bm z$ in one direction $j$ only, $G_{jj}(t)=G_{jj}(0)+2t\,\bm z^T\bm b_j$, which is non-negative for every real $t$ only if $\bm z^T\bm b_j=0$.  Each column $\bm b_j$ is therefore orthogonal to $\ker W$, that is, in $\operatorname{Ran}W$, the normal equations $W\bm\eta^\star=-\bm b$ are solvable, and completing the square gives
\begin{equation}
G_e=G_\star+\Gamma(\bm\eta-\bm\eta^\star),\qquad
G_\star=G-\bm b^TW^+\bm b,\qquad
\Gamma(\bm\zeta)_{ij}=\sum_{\alpha\beta}\zeta_{\alpha i}W_{\alpha\beta}\zeta_{\beta j},\qquad\bm\eta^\star=-W^+\bm b .
\label{S:gram}
\end{equation}
For real $\bm c$, $\bm c^T\Gamma(\bm\zeta)\bm c$ is the Brillouin-zone integral of the variance of $\sum_jc_j\zeta_{\alpha j}$ in $w_\alpha$, so $G_e\succeq G_\star$ for every embedding.  The $x$ and $y$ displacements enter through the same form $W$ and are not coupled by the linear term; this is why minimizing the trace \cite{Huhtinen2022} already gives the Loewner minimum.  Since $C\neq0$ forces $G_\star\succ0$ by Eq.~\eqref{S:chain},
\begin{equation}
\det G_e=\det G_\star\det\bigl(I+G_\star^{-1/2}\Gamma G_\star^{-1/2}\bigr)\ge\det G_\star,
\label{S:detmin}
\end{equation}
with equality if and only if $\Gamma(\bm\eta-\bm\eta^\star)=0$.

\subsection*{S3.2 Fisher-covariance form}
Where $w_\alpha\neq0$ write $u_\alpha=\sqrt{w_\alpha}\,\e^{\ii\Phi_\alpha}$.  Then
\begin{equation}
g_{ij}=\sum_\alpha\partial_i\sqrt{w_\alpha}\,\partial_j\sqrt{w_\alpha}+\mathrm{Cov}_w(\partial_i\Phi,\partial_j\Phi),\qquad
\beta_{\alpha j}=-w_\alpha\Bigl(\partial_j\Phi_\alpha-\sum_\beta w_\beta\partial_j\Phi_\beta\Bigr).
\label{S:fisher}
\end{equation}
The first term, a quarter of the Fisher metric of the orbital distribution, does not depend on the embedding; an embedding shifts the phase gradients $\partial_j\Phi_\alpha$ by $-\eta_{\alpha j}$ inside the covariance.  The natural embedding is the least-squares fit of constant orbital positions to the momentum-dependent phase gradients, and $G_\star$ is the Fisher term plus the residual of that fit.

\subsection*{S3.3 Site-rigid restriction and the splitting defect}
Let $\Pi$ be the $N\times N_{\rm s}$ incidence matrix of orbitals and sites.  Site-rigid embeddings are $\bm\eta=\Pi\bm\rho$, and Eq.~\eqref{S:Gquad} becomes a form of the same type with $W\to\Pi^TW\Pi$ and $\bm b\to\Pi^T\bm b$.  Its minimizer is $\bm\eta^\Vcal=\Pi\bm\rho^\star$, $\bm\rho^\star=-(\Pi^TW\Pi)^+\Pi^T\bm b$, the projection of $\bm\eta^\star$ onto $\Vcal$ orthogonal with respect to $W$, and
\begin{equation}
G_\Vcal=G_\star+\Sigma,\qquad\Sigma=\Gamma(\bm\eta^\Vcal-\bm\eta^\star)\succeq0 .
\label{S:sigma}
\end{equation}
$\Sigma$ vanishes if and only if some natural embedding is site rigid; its rank is at most $\min(2,N-N_{\rm s})$.  When every site hosts one orbital, $\Pi$ is the identity and $\Sigma=0$.  A rotation about a site that acts on its orbitals by phases only keeps them together, because their relative displacements must then be rotation invariant.

\subsection*{S3.4 Two orbitals}
For $N=2$, $W=w_{12}\begin{psmallmatrix}1&-1\\-1&1\end{psmallmatrix}$ with $w_{12}=\int_\BZ w_1w_2\,\dd^2k$, and the orbital rows of $\bm b$ obey $\bm b_{(2)}=-\bm b_{(1)}$.  In terms of the relative displacement $\bm d=\bm\eta_2-\bm\eta_1$,
\begin{equation}
G_e=G_\star+w_{12}(\bm d-\bm d_\star)(\bm d-\bm d_\star)^T,\qquad\bm d_\star=\bm b_{(1)}/w_{12},\qquad
\det G_e=\det G_\star\bigl[1+w_{12}(\bm d-\bm d_\star)^TG_\star^{-1}(\bm d-\bm d_\star)\bigr].
\label{S:twoorb}
\end{equation}
If the two orbitals share a site and the reference is that site, $\Sigma=w_{12}\bm d_\star\bm d_\star^T$.

\subsection*{S3.5 Orbital-resolved vortexability}
$\Torb=0$ means $\det\bar Q_{e_\star}=0$, with $e_\star$ the natural embedding $\bm\eta=\bm\eta^\star$.  As an integral of positive-semidefinite matrices, $\bar Q_{e_\star}$ then has a constant null vector $\bm c\in\mathbb C^2$ shared by all $Q_{e_\star}(\bm k)$, so $(1-P_{e_\star})\sum_jc_j\partial_ju_{e_\star}=0$.  With $\partial_j\psi_{\bm k}=\ii\hat r_j\psi_{\bm k}+\e^{\ii\bm k\cdot\hat{\bm r}}\partial_ju_{\bm k}$, where $\hat{\bm r}$ is the position operator of the embedding $e_\star$, this is
\begin{equation}
(1-\mathcal P)\,\hat z\,\mathcal P=0,\qquad\hat z=\sum_{\bm R\alpha}\bm c\cdot(\bm R+\bm\delta_\alpha+\bm\eta^\star_\alpha)\,|\bm R\alpha\rangle\langle\bm R\alpha|,
\end{equation}
with $\mathcal P$ the projector on the band.  If $\Sigma=0$, $\hat z$ takes one value per site and is a zero of the site-rigid functional, $\TLVPcl=0$, and actual vortexability follows when the site-rigid optimum is attained by an admissible coordinate.  If $\Sigma\neq0$, $\hat z$ distinguishes orbitals of one site: it maps the band into itself, but the Jastrow factor built from it does not vanish when two particles occupy different orbitals of that site.  This is the tight-binding counterpart of the layer-dependent vortex operator that LVP propose for strained chiral twisted bilayer graphene \cite{LVP2023}.  In that model, for an exactly flat band, no such operator could map the band into itself (Sec.~S6.4); numerically $\Torb>0$ and $\Sigma\neq0$ there.  Altogether $\TLVP=0\Rightarrow\TLVPcl=0\Rightarrow\Torb=0$.

\subsection*{S3.6 The least integrated metric of any real-space realization}
A change of the lattice shape at fixed cell area acts on the integrated metric as $G\to\mathsf MG\mathsf M^T$ with $\det\mathsf M=1$, and the same relative embedding minimizes every diagonal component for every choice of lattice vectors \cite{WRS2026}.  Since $\det(\mathsf MG\mathsf M^T)=\det G$ and $\min_{\det\mathsf M=1}\tr(\mathsf MG\mathsf M^T)=2\sqrt{\det G}$ for $G\succ0$, Loewner minimality gives
\begin{equation}
\min_{\text{positions, lattice shape}}\int_\BZ\tr g\,\dd^2k=2\sqrt{\det G_\star}=\bigl(4\pi^2C^2+\Torb^2\bigr)^{1/2}.
\end{equation}
The topological floor is therefore reached exactly when some real-space realization of the band is ideal.

\subsection*{S3.7 Numerical checks}
\texttt{checks\_gram\_identity.py} draws a random four-orbital Bloch Hamiltonian with hopping to the eight nearest cells and takes its lowest band (minimum gap $2.4$, $C=-2$), with orbitals 2 and 3 declared to share a site.  On $60^2$ and $90^2$ grids, where Eq.~\eqref{S:gram} is algebraic and must hold exactly, the residual over 2000 random embeddings is at most $8\times10^{-12}$, $G_e-G_\star$ stays positive semidefinite, and $G_\Vcal=G_\star+\Sigma$ to $10^{-13}$ with eigenvalues of $\Sigma$ equal to $0$ and $0.959$, as the rank bound requires.  For this band $1+(\Torb/2\pi|C|)^2=11.337$ and $1+(\TLVPcl/2\pi|C|)^2=11.825$ at both grids.  \texttt{check\_qwzx\_rank2.py} repeats the identity for the composite of the two lowest bands of a random four-orbital model ($C=-1$, gap $0.21$ to the third band), with $W_{\alpha\beta}=\int(P_{\alpha\alpha}\delta_{\alpha\beta}-|P_{\alpha\beta}|^2)$: on $256^2$ and $384^2$ grids, the residual over random embeddings is below $5\times10^{-9}$ and $G_e-G_\star\succeq0$; the same script recomputes three rows of Table~\ref{tab:qwzgeom} ($M=-1.8$, $-1$, $-0.6$) at $200^2$ and $400^2$.

\section*{S4. The exact zero-temperature stiffness}

\subsection*{S4.1 Model and mean-field prefactor}
The spin-up electrons occupy an isolated, exactly flat band with projector $P(\bm k)$ on $N$ orbitals; the spin-down band is its time-reversed partner, $P_\downarrow(\bm k)=\overline{P(-\bm k)}$.  The interaction is the on-site, orbital-diagonal attraction $-|U|\sum_{\bm R\alpha}n_{\bm R\alpha\uparrow}n_{\bm R\alpha\downarrow}$, and remote bands are projected out.  Uniform pairing means that the Brillouin-zone-averaged weight $\bar w_\alpha=(2\pi)^{-2}\int_\BZ|u_\alpha|^2\,\dd^2k$ (unit cell area one) takes the same value $n_\phi$ on every orbital that carries the band; orbitals with $\bar w_\alpha=0$ play no role.  For $N_{\rm orb}$ active orbitals, $n_\phi=1/N_{\rm orb}$.  With the band at zero energy, the flat-band gap and number equations give the quasiparticle energy $E=|U|n_\phi/2$ and the pairing amplitude $\Delta=|U|n_\phi\sqrt{\nu(1-\nu)}$ at filling $\nu$.  For pair momentum $2\bm q$ the quasiparticle energy is $E_{\bm k}(\bm q)=[\mu^2+\Delta^2|\langle u_{\bm k+\bm q}|u_{\bm k-\bm q}\rangle|^2]^{1/2}=E-(2\Delta^2/E)\,g_{ij}q_iq_j+O(q^3)$, which gives the unrelaxed Hessian $\kappa_0G_e$ in the embedding $e$ \cite{PT2015}, and relaxation of the orbital phases replaces $G_e$ by $G_\star$ \cite{Huhtinen2022,TP2024} (Sec.~S4.5), Eq.~\eqref{S:kappa0}.  Since $E=|U|n_\phi/2$ does not depend on $\nu$, the prefactor $2\Delta^2/E=4|U|n_\phi\nu(1-\nu)$ holds at every filling.

The weight per active orbital matters for comparisons between bands.  Splitting every orbital into two co-located copies with amplitudes $1/\sqrt2$ halves every projected density and doubles the number of orbitals; the projected Hamiltonian, and with it the whole spectrum at every twist, is halved, while $G_\star$ does not change (numerical check (C) of \texttt{chain\_checks.py}, residual below $10^{-14}$).  At fixed $|U|$ a band spread over more orbitals is therefore less stiff, which is why the floor $\det\Ds^{\rm floor}=\kappa_0^2\pi^2C^2$ refers to fixed $\kappa_0$, that is, fixed filling and pairing amplitude.

\subsection*{S4.2 Exactness}
Two independent proofs show that the stiffness of nondegenerate quantum-geometric-nesting (QGN) superconductors equals its BCS value \cite{HKH2026,WLY2026}, and the time-reversed flat band with Hubbard attraction and uniform pairing is the simplest member of that family \cite{HHBK2024,HKH2026}; for a $C=\pm1$ band the rigorous bootstrap lower bound of Ref.~\cite{GHK2026} matches the BCS value numerically.  We give a short argument for our setting, which reproduces the finite-mesh form of Ref.~\cite{WLY2026} [their Eq.~(S105)].

Write $\bar c_{\bm R\alpha\sigma}$ for the projected annihilation operators on $N_{\rm c}$ cells, $\bar n_{\bm R\alpha\sigma}=\bar c^\dagger_{\bm R\alpha\sigma}\bar c_{\bm R\alpha\sigma}$, and $S_{\bm R\alpha}=\bar n_{\bm R\alpha\uparrow}-\bar n_{\bm R\alpha\downarrow}$.  Each $\bar c_{\bm R\alpha\sigma}$ is a linear combination of fermion annihilation operators, so $\bar c_{\bm R\alpha\sigma}^2=0$, and $\{\bar c_{\bm R\alpha\sigma},\bar c^\dagger_{\bm R\alpha\sigma}\}=\bar w_\alpha$, the on-site element of the band projector; hence $\bar n_{\bm R\alpha\sigma}^2=\bar w_\alpha\bar n_{\bm R\alpha\sigma}$ and, under uniform pairing, with $\hat N=\sum_{\bm R\alpha\sigma}\bar n_{\bm R\alpha\sigma}$ the particle number,
\begin{equation}
-|U|\sum_{\bm R\alpha}\bar n_{\bm R\alpha\uparrow}\bar n_{\bm R\alpha\downarrow}=\frac{|U|}{2}\sum_{\bm R\alpha}S_{\bm R\alpha}^2-\frac{|U|n_\phi}{2}\hat N .
\label{S:qgn}
\end{equation}
Let $\gamma_{\bm k\sigma}$ annihilate the band states, with the down-spin state at $\bm k$ the time reverse of the up-spin state at $-\bm k$, and let $\eta^+=\sum_{\bm k}\gamma^\dagger_{\bm k\uparrow}\gamma^\dagger_{-\bm k\downarrow}$, $\eta^-=(\eta^+)^\dagger$, $\eta^z=\frac12(\hat N-N_{\rm c})$.  A direct computation gives $[\bar n_{\bm R\alpha\uparrow},\eta^+]=\bar c^\dagger_{\bm R\alpha\uparrow}\bar c^\dagger_{\bm R\alpha\downarrow}=[\bar n_{\bm R\alpha\downarrow},\eta^+]$, so every $S_{\bm R\alpha}$ commutes with the $\eta$-pairing SU(2) and is an SU(2) scalar \cite{Yang1989,Tovmasyan2016,HHBK2024}.  The Hamiltonian $\mathcal H=\frac{|U|}{2}\sum S^2$ is positive semidefinite and annihilates the vacuum, so the states $|s,m\rangle\propto(\eta^+)^{N_p}|0\rangle$, with $s=N_{\rm c}/2$ and $m=N_p-N_{\rm c}/2$, are zero-energy ground states in every sector of $N_p$ pairs.

Twist the model by a uniform vector potential, $\bm k\to\bm k+\bm A$ for both species.  We study the twisted QGN Hamiltonian $\mathcal H(\bm A)=\frac{|U|}{2N_{\rm c}}\sum_{\alpha\bm Q}S_{\alpha\bm Q}(\bm A)S_{\alpha\bm Q}(\bm A)^\dagger$, with $S_{\alpha\bm Q}$ the Fourier components at momentum transfer $\bm Q$; $S(\bm A)$ is a number-conserving fermion bilinear for every $\bm A$.  The twisted projected Hubbard model differs from $\mathcal H(\bm A)$ by the one-body term $-\frac{|U|}{2}\sum_{\bm R\alpha\sigma}\bar w_{\alpha\sigma}(\bm A)\,\bar n_{\bm R\alpha\sigma}$, whose grid weights $\bar w_{\alpha\sigma}(\bm A)$ depend on $\bm A$ only through aliasing; the difference is exponentially small in the linear size for an analytic band and vanishes in the thermodynamic limit.  Assume that $|s,m\rangle$ is the only ground state in its sector.  Since $S^\dagger|s,m\rangle=0$, second-order perturbation theory gives
\begin{equation}
E''_{ij}(N_p)=\langle\mathcal Y_{ij}\rangle-2\,\mathrm{Re}\,\langle\mathcal X_i^\dagger\mathcal H^+\mathcal X_j\rangle,\qquad
\mathcal X_i=\frac{|U|}{2N_{\rm c}}\sum_{\alpha\bm Q}S_{\alpha\bm Q}\,\partial_iS^\dagger_{\alpha\bm Q},\qquad
\mathcal Y_{ij}=\frac{|U|}{2N_{\rm c}}\sum_{\alpha\bm Q}\bigl(\partial_iS_{\alpha\bm Q}\,\partial_jS^\dagger_{\alpha\bm Q}+\partial_jS_{\alpha\bm Q}\,\partial_iS^\dagger_{\alpha\bm Q}\bigr),
\label{S:pt2}
\end{equation}
with expectation values in $|s,m\rangle$, derivatives with respect to $\bm A$, and $\mathcal H^+$ the inverse of $\mathcal H$ on the complement of the ground state.  A number-conserving bilinear $\mathcal O$ obeys $[\eta^+,[\eta^+,\mathcal O]]=0$, so it is a sum of SU(2) tensors of rank zero and one.  $S$ and $\mathcal H^+$ are scalars and $\partial_iS$ has rank at most one, so $\mathcal X_i$ has rank at most one and both terms in Eq.~\eqref{S:pt2} are expectation values of tensors of rank at most two in the multiplet $|s,m\rangle$.  By the Wigner--Eckart theorem $E''(N_p)$ is a polynomial of degree at most two in $m$.  It vanishes at $N_p=0$, so
\begin{equation}
E''(N_p)=E''(1)\,\frac{N_p(N_{\rm c}-N_p)}{N_{\rm c}-1}+E''(N_{\rm c})\,\frac{N_p(N_p-1)}{N_{\rm c}(N_{\rm c}-1)} .
\label{S:quadlaw}
\end{equation}
The full-band term $E''(N_{\rm c})$ comes from grid aliasing of the orbital weights at finite twist: for the filled band $E(\bm A)=\frac{|U|}{2}N_{\rm c}\sum_\alpha[\bar w_{\alpha\uparrow}(\bm A)-\bar w_{\alpha\downarrow}(\bm A)]^2$, which is exponentially small in the linear size for an analytic band.  The one-pair curvature is a two-body quantity.  One pair at twist $\bm A$ is a pair of momentum $2\bm A$ in the untwisted band, so its energy is $E_0(2\bm A)$ of Sec.~S4.4 and $E''(1)=4|U|X$, with $X=2n_\phi G_\star/4\pi^2$ derived there: $E''(1)=8|U|n_\phi\gbar$ \cite{HA2026,HKH2026}.  The stiffness per cell at filling $\nu=N_p/N_{\rm c}$ is therefore $\Ds=\lim E''(N_p)/N_{\rm c}=8|U|n_\phi\,\gbar\,\nu(1-\nu)=\kappa_0G_\star$.  This reproduces the theorem of Refs.~\cite{HKH2026,WLY2026}; in our setting the argument needs only nondegeneracy and the two-body limit.

\subsection*{S4.3 Exact diagonalization}
\texttt{ed\_qgn.py} diagonalizes the band-projected QGN Hamiltonian $\mathcal H(\bm A)$ for three bands of this work on $3\times3$ and $3\times4$ tori, in every sector of $N_p$ pairs on the $3\times3$ torus and for $N_p=1,2,3,11,12$ on the $3\times4$ torus (the script prints $N_p$ as $p$ and $N_{\rm c}$ as $M$).  On the $3\times4$ torus the grid-averaged orbital weights of the $\pi$-flux band are $0.512$ and $0.488$, so there the QGN Hamiltonian differs from the projected Hubbard model by a one-body term, and that row tests the QGN argument rather than uniform pairing.  The other two bands have equal grid weights on both tori.  Curvatures come from energies at $\pm h\hat{\bm e}$ for three directions with Richardson extrapolation in the finite-difference step $h$.  In every sector $E_0(\bm 0)=0$ to $2\times10^{-15}$, and in every sector with $N_p<N_{\rm c}$ there is a positive gap to the next state, so the $\eta$-pairing state is nondegenerate on these tori, and the curvature tensor obeys Eq.~\eqref{S:quadlaw} to the accuracy of the finite differences (Table~\ref{tab:ed}).  The one-pair curvature on larger tori converges to $8|U|n_\phi\gbar$ as a full tensor (Table~\ref{tab:twobody}).

\begin{table}[h]
\caption{Exact diagonalization of the band-projected QGN model ($|U|=1$).  Gap: to the first excited state, the same in every sector with $N_p<N_{\rm c}$ (the filled band has a single state).  Deviation: largest difference between the computed curvature tensor and Eq.~\eqref{S:quadlaw}.  $E''_{xx}(N_{\rm c})$: full-band aliasing term.  Numerical (\texttt{ed\_qgn.py}).}
\label{tab:ed}
\centering
\begin{tabular}{lccccc}
\toprule
model & torus & sectors $N_p$ & gap$/|U|$ & max deviation & $E''_{xx}(N_{\rm c})$\\
\midrule
QWZ, $M=-1$ & $3\times3$ & 1 to 9 & 0.103433 & $1.1\times10^{-9}$ & 0.1002\\
 & $3\times4$ & 1, 2, 3, 11, 12 & 0.062993 & $1.4\times10^{-8}$ & 0.1274\\
theta, $d=0.5$ & $3\times3$ & 1 to 9 & 0.130018 & $4.6\times10^{-9}$ & $1.5\times10^{-4}$\\
 & $3\times4$ & 1, 2, 3, 11, 12 & 0.081946 & $3.0\times10^{-9}$ & $1.9\times10^{-4}$\\
$\pi$ flux & $3\times3$ & 1 to 9 & 0.144565 & $4.4\times10^{-10}$ & 0\\
 & $3\times4$ & 1, 2, 3, 11, 12 & 0.089113 & $1.3\times10^{-9}$ & 0\\
\bottomrule
\end{tabular}
\end{table}

\begin{table}[h]
\caption{One-pair curvature $E''(1)$ on $L\times L$ tori against $4|U|\gbar$ ($n_\phi=1/2$): ratios of the diagonal components (off-diagonal components vanish).  The QWZ band at $M=-1$ is anisotropic, $\bar g_{\star,yy}/\bar g_{\star,xx}=0.6005$, and the ratio converges for both components.  Numerical (\texttt{two\_body.py}).}
\label{tab:twobody}
\centering
\begin{tabular}{lccccc}
\toprule
model & $L=8$ & $L=12$ & $L=16$ & $L=24$ & $L=32$\\
\midrule
QWZ, $M=-1$, $xx$ / $yy$ & 1.0125 / 1.0224 & 1.00093 / 1.00164 & 1.000066 / 1.000116 & 1.000000 / 1.000001 & 1.000000 / 1.000000\\
theta, $d=0.5$ & & 1.000000 & & 1.000000 & 1.000000\\
$\pi$ flux & & 1.000018 & & 1.000000 & 1.000000\\
\bottomrule
\end{tabular}
\end{table}

A one-dimensional check isolates the ingredients (\texttt{chain\_checks.py}).  A two-orbital chain band with $|u_1|^2=\frac12[1+0.5\sin k+0.3\sin2k]$ keeps its grid-averaged orbital weights equal to $1/2$ at every twist, so uniform pairing holds exactly on finite grids, and its natural orbital displacement is nonzero ($-0.0813$).  Exact diagonalization on $L=6$ and $8$ cells reproduces $E''(N_p)/E''(1)=N_p(L-N_p)/(L-1)$ for every pair number to $6\times10^{-9}$, the accuracy of the finite differences, and the one-pair kernel gives $E''(1)=0.71541754$ at $L=50$, $100$, $200$, equal to $8|U|n_\phi\gbar$, whereas the metric at the physical positions would give $0.72091$.

\subsection*{S4.4 Exact Cooper-pair mass}
Herzog-Arbeitman \emph{et al.}\ show that, under uniform pairing, the condensing Cooper pair of the projected attractive Hubbard model has the exact dispersion $E_0(\bm p)=|U|[\epsilon_0(\bm 0)-\epsilon_0(\bm p)]$, with $\epsilon_0$ the top eigenvalue of the binding matrix $\mathcal B_{\alpha\beta}(\bm p)=N_{\rm c}^{-1}\sum_{\bm k}P_{\alpha\beta}(\bm k+\bm p)P_{\beta\alpha}(\bm k)$ on $N_{\rm c}$ cells, and identify its curvature with the minimal metric \cite{HA2026}.  The link between the mass of a two-body bound state in a flat band and the quantum metric was found earlier \cite{TLP2018,Iskin2021,Iskin2022}.  We record the short derivation in our notation.  An embedding conjugates $\mathcal B(\bm p)$ by the constant diagonal unitary $\mathrm{diag}(\e^{-\ii\bm p\cdot\bm\eta_\alpha})$, so its eigenvalues, and the Hessian $X=-\partial^2\epsilon_0|_{\bm 0}$, do not depend on the embedding, and we may work at the natural embedding.  $\mathcal B(\bm p)$ is Hermitian, and at $\bm p=\bm 0$ it is entrywise non-negative with row sums $N_{\rm c}^{-1}\sum_{\bm k}P_{\alpha\alpha}(\bm k)=\bar w_\alpha$.  Under uniform pairing the vector $\bm a$ with entries $N_{\rm orb}^{-1/2}$ on the active orbitals is therefore an eigenvector with the largest eigenvalue, $n_\phi$, which is nondegenerate whenever the one-pair ground state is, as part (ii) of the theorem assumes.  Since $(\partial_iP\,P)_{\alpha\alpha}=u_\alpha^*((1-P)\partial_iu)_\alpha$, the vector $\partial_i\mathcal B\,\bm a$ has entries $N_{\rm orb}^{-1/2}N_{\rm c}^{-1}\sum_{\bm k}u_\alpha^*((1-P)\partial_iu)_\alpha$, whose real part is the grid sum of the total derivative $\frac12\partial_iw_\alpha$ and whose imaginary part is proportional to $b_{\alpha i}$.  Both vanish at the natural embedding (the first up to aliasing), so the first-order shift of $\epsilon_0$ and the second-order term through the other eigenvectors vanish, and
\begin{equation}
X_{ij}=-\bm a^T\partial_i\partial_j\mathcal B\,\bm a=-\frac{1}{N_{\rm orb}N_{\rm c}}\sum_{\bm k}\Tr\bigl[\partial_i\partial_jP\,P\bigr]=\frac{2}{N_{\rm orb}N_{\rm c}}\sum_{\bm k}g^\star_{ij}(\bm k)\ \longrightarrow\ \frac{2n_\phi}{4\pi^2}\,G_{\star,ij},
\label{S:pairmass}
\end{equation}
where $g^\star$ is the metric at the natural embedding and $\Tr[\partial_i\partial_jP\,P]=-\Tr[\partial_iP\,\partial_jP]=-2g_{ij}$ follows from differentiating $\Tr P^2=1$ twice.  The exact inverse-mass tensor of the pair is $|U|X$, and $(2\pi^2/n_\phi)^2\det X/\pi^2C^2=1+(\Torb/2\pi|C|)^2$.  Table~\ref{tab:pairmass} confirms $X=G_\star/4\pi^2$ ($n_\phi=1/2$) as a full tensor, including the anisotropic QWZ case.  For the theta band in its one-site realization the exact pair mass sits on the floor although the band is not vortexable.

\begin{table}[h]
\caption{Exact two-body pair Hessian $X$ against the natural metric ($n_\phi=1/2$, so $X=G_\star/4\pi^2$ is predicted).  $XG_\star^{-1}=0.0253303\,I=(4\pi^2)^{-1}I$ in every case, while $XG^{-1}$ in the one-site frame is not a multiple of the identity.  Grids $96^2$ and $192^2$ agree in every case.  Numerical (\texttt{pair\_mass.py}).}
\label{tab:pairmass}
\centering
\begin{tabular}{lccccc}
\toprule
model & grid & $X$ (diagonal) & $XG_\star^{-1}$ & $XG^{-1}$, one site & $(4\pi^2)^2\det X/\pi^2$ ; $\det G_\star/\pi^2$\\
\midrule
QWZ, $M=-1$ & $96^2$, $192^2$ & (0.1134077, 0.0680968) & $0.0253303\,I$ & diag(0.02533, 0.01521) & 1.219520 ; 1.219520\\
QWZ, $M=-1.5$ & $96^2$, $192^2$ & (0.0970771, 0.0735142) & $0.0253303\,I$ & diag(0.02533, 0.01918) & 1.126959 ; 1.126959\\
theta, $d=0.5$ & $96^2$, $192^2$ & (0.0795775, 0.0795775) & $0.0253303\,I$ & diag(0.02533, 0.0153) & 1.000000 ; 1.000000\\
\bottomrule
\end{tabular}
\end{table}

\subsection*{S4.5 Free-energy Hessian and phase response}
Let $\mathcal F_e(\bm q,\xi)$ be the mean-field grand potential at phase twist $\bm q$ in the embedding $e$, with $\xi$ the internal variables (amplitudes and relative phases $\theta_\alpha$ of the orbital order parameters, chemical potential) after the global phase is fixed, for the ansatz $\Delta_{\bm R\alpha}=|\Delta_\alpha|\e^{\ii\theta_\alpha}\e^{2\ii\bm q\cdot(\bm R+\bm\delta_\alpha+\bm\eta_\alpha)}$.  With $A_e=\partial^2_{\bm q\bm q}\mathcal F_e$, $B_e=\partial^2_{\bm q\xi}\mathcal F_e$, $H_e=\partial^2_{\xi\xi}\mathcal F_e$ at the saddle, differentiating $\partial_\xi\mathcal F_e(\bm q,\xi(\bm q))=0$ gives $\dd\xi/\dd\bm q=-H_e^{-1}B_e^T$ and the relaxed Hessian
\begin{equation}
F_{qq}=A_e-B_eH_e^{-1}B_e^T,
\label{S:schur}
\end{equation}
the Schur complement of $H_e$.  Time reversal makes the amplitudes, the chemical potential, and the Hartree shifts even in $\bm q$, and at $\bm q=0$ the grand potential is even in the relative phases about the real saddle, so only the phase block of $H_e$ and the phase columns of $B_e$ enter Eq.~\eqref{S:schur}.

In the ansatz, the phase shift $\theta_\alpha\to\theta_\alpha+2\bm q\cdot\bm\zeta_\alpha$ is the same as the embedding shift $\bm\eta_\alpha\to\bm\eta_\alpha+\bm\zeta_\alpha$, so
\begin{equation}
\mathcal F_e(\bm q,\theta+2\bm\zeta\bm q)=\mathcal F_{e+\zeta}(\bm q,\theta)
\label{S:shift}
\end{equation}
with the other internal variables held fixed.  In the flat uniform-pairing limit the frozen Hessian is $A_{e'}=\kappa_0G_{e'}$ in every embedding $e'$ (Sec.~S4.1).  Expand both sides of Eq.~\eqref{S:shift} at $\theta=0$ to second order in $\bm q$: the left side gives $\frac12\bm q^T[A_e+2(B_e\bm\zeta+\bm\zeta^TB_e^T)+4\bm\zeta^TH_e\bm\zeta]\bm q$, with $B_e$ and $H_e$ restricted to the phases, and the right side gives $\frac12\kappa_0\bm q^TG_{e+\zeta}\bm q$ with $G_{e+\zeta}=G_e+\bm\zeta^T\bm b_e+\bm b_e^T\bm\zeta+\bm\zeta^TW\bm\zeta$ and $\bm b_e=\bm b+W\bm\eta$, by Eq.~\eqref{S:Gquad}.  Since $\bm\zeta$ is arbitrary,
\begin{equation}
B_e=\tfrac12\kappa_0\bm b_e^T,\qquad H_e=\tfrac14\kappa_0W,\qquad
F_{qq}=\kappa_0\bigl(G_e-\bm b_e^TW^+\bm b_e\bigr)=\kappa_0G_\star,\qquad
\frac{\dd\theta}{\dd\bm q}=-2W^+\bm b_e=2(\bm\eta^\star-\bm\eta),
\label{S:phase}
\end{equation}
the last up to a constant common to all orbitals, with $H_e^{-1}$ read as the pseudoinverse on the relative phases.  We used $\bm b=-W\bm\eta^\star$, so that $\bm b_e=W(\bm\eta-\bm\eta^\star)$ and $\bm b_e^TW^+\bm b_e=\Gamma(\bm\eta-\bm\eta^\star)$.  At the natural embedding $\bm b_{e_\star}=0$, so $B_{e_\star}=0$ and the first-order phase response vanishes there, as found in Ref.~\cite{Huhtinen2022}; in any embedding $B_eH_e^{-1}B_e^T=\kappa_0\Gamma(\bm\eta-\bm\eta^\star)$.  The reduction of the stiffness by phase relaxation is exactly the Gram term, and in mean-field theory the natural embedding is a property of the paired state, read off from its relative-phase response.  Both statements are tested in Secs.~S5.1 and S5.6.

\subsection*{S4.6 The meaning of ``least stiff''}
The ordering of the Letter compares bands at fixed Chern number, filling, and pairing amplitude, equivalently at fixed on-site attraction and weight per active orbital, and it belongs to the on-site attraction.  Three remarks delimit it.  (i) A trivial band can have a smaller stiffness than any Chern band; the floor is a statement within fixed $|C|$.  (ii) Other QGN interactions change the stiffness of a given band: a nearest-neighbor $S^zS^z$ coupling $J_{zz}$ keeps the $\pi$-flux model frustration free for $2|J_{zz}|\le|U|$ but changes its stiffness \cite{GHK2026}.  (iii) For interlayer exciton condensates with gate-screened Coulomb interactions, pure Landau levels are stiffer than mixtures with the same quantum geometric tensor \cite{Zou2026}.  Under component-diagonal contact attraction with uniform pairing, the pure $n=1$ level and the equal mixture of the $n=0$ and $n=2$ levels both have $R=9$ (Sec.~S5.5).  At equal attraction the mixture, with weight $1/2$ per component, is half as stiff; at equal pairing amplitude the two are equally stiff.  These continuum statements are mean field; they become exact if the argument of Sec.~S4.2 carries over from the lattice, which we have not shown, and they are predictions, not computations.

\section*{S5. Examples}

\subsection*{S5.1 The theta band}
Exactly flat lattice bands built by sampling the lowest Landau level at lattice sites go back to Kapit and Mueller \cite{KM2010} and to Dong and Mueller \cite{DM2020}; Wang \emph{et al.}\ gave an exact Landau-level description of ideal flat bands, with exact fractional Chern insulator ground states of short-range repulsion \cite{Wang2021}; and Li and Wu, Zhao \emph{et al.}, and Wang, Raca, and Simon construct ideal bands with ordinary translation symmetry and several orbitals per unit cell \cite{LiWu2026,Zhao2026,WRS2026}.  We use the square-lattice instance with one flux quantum per cell and two sampling points a distance $d$ apart along $y$, written in the one-site periodic gauge as
\begin{equation}
v(\bm k)=\Bigl(\vartheta_1\bigl(z+\tfrac d2\big|\tau\bigr),\ \e^{-\ii k_yd}\,\vartheta_1\bigl(z-\tfrac d2\big|\tau\bigr)\Bigr),\qquad z=\frac{k_x+\ii k_y}{2\pi},\qquad\tau=\ii,
\label{S:theta}
\end{equation}
normalized to $u=v/|v|$ (\texttt{theta\_model.py}), with $\vartheta_1(z|\tau)=2\sum_{n\ge0}(-1)^n\e^{\ii\pi\tau(n+1/2)^2}\sin[(2n+1)\pi z]$, so that $\vartheta_1(z+1|\tau)=-\vartheta_1(z|\tau)$ and $\vartheta_1(z+\tau|\tau)=-\e^{-\ii\pi\tau-2\pi\ii z}\vartheta_1(z|\tau)$.  Under $\bm k\to\bm k+\bm K$, with $\bm K$ a reciprocal vector, the vector $v$ changes only by an overall factor, so $P$ is periodic.  In the embedding that displaces orbital 2 by $(0,-d)$ relative to orbital 1, the components become $\vartheta_1(z\pm d/2)$, holomorphic in $z$: the band is ideal there, $G_\star=\pi I$ and $\Torb=0$, and the natural displacement is $\bm d_\star=(0,-d)$.  The symmetry $P(-\bm k)=\sigma_xP(\bm k)\sigma_x$ makes the integrated orbital weights equal, so uniform pairing holds.

The same Bloch states describe two physical systems.  With the orbitals on two sites at the sampling points, the physical embedding is natural, $\Sigma=0$, and the band is vortexable.  With both orbitals on one site the only site-rigid embedding is the common one (up to translation), so $\TLVPcl=\TLVP$ is given by the one-site metric, $\det G/\pi^2=1+w_{12}d^2/\pi$, while $\Sigma=w_{12}\bm d_\star\bm d_\star^T\neq0$ (Table~\ref{tab:theta}).  Its stiffness is that of the two-site system, because the paired state does not know where the orbitals sit.

\begin{table}[h]
\caption{Geometry of the theta band, identical at $96^2$ and $192^2$ (\texttt{theta\_model.py}).  $C=-1$ and the orbital weights are $0.5/0.5$ in all cases.  Numerical.}
\label{tab:theta}
\centering
\begin{tabular}{cccccc}
\toprule
$d$ & $G_\star/\pi$ & $\bm d_\star$ & $w_{12}$ & one site: $\det G/\pi^2=1+w_{12}d^2/\pi$ & one site: $\TLVPcl/2\pi=\TLVP/2\pi$\\
\midrule
0.1 & $I$ & $(0,-0.1)$ & 9.472764 & 1.03015274 & 0.173645\\
0.25 & $I$ & $(0,-0.25)$ & 8.748990 & 1.17405563 & 0.417200\\
0.5 & $I$ & $(0,-0.5)$ & 8.243101 & 1.65596511 & 0.809917\\
\bottomrule
\end{tabular}
\end{table}

\paragraph{Self-consistent BdG.}  \texttt{theta\_bdg\_scan.py} solves the mean-field problem of Sec.~S5.6 with the QWZ projector replaced by that of Eq.~\eqref{S:theta}: spin-up $h(\bm k)=E_g[1-P(\bm k)]$, its time-reversed partner, on-site attraction $|U|=1$, $\mu=0$, twist step $0.005$.  Table~\ref{tab:thetabdg} shows that the relaxed stiffness sits on the floor up to a residual that falls by about a factor of four when $E_g$ doubles, that the frozen-phase stiffness is that of the one-site frame, and that the relative-phase response measures the sampling distance, $\frac12\dd\theta/\dd q_y=-d$.  For $d=0.1$ the band varies faster in $\bm k$: a twist step of $0.02$ leaves a finite-difference error of $1.7\times10^{-5}$ in the determinant ratio, while the steps $0.005$ and $0.0025$ agree to $2\times10^{-7}$ (and $0.01$ to $1.2\times10^{-6}$); and a $48^2$ grid is too coarse ($10^{-5}$), while $96^2$, $144^2$, and $192^2$ agree to the digits shown.

\begin{table}[h]
\caption{Self-consistent BdG for the flattened theta band, half filling, $|\Delta|/|U|=0.25078$ at $E_g=80|U|$, twist step $0.005$.  Ratios are $\det(\cdot)/(\kappa_H\pi)^2$; the theorem gives $1$ for the relaxed stiffness, and the column lists the residual $1-{}$relaxed, which the twist step leaves uncertain by less than $10^{-6}$.  The Schur evaluation agrees with the relaxed stiffness to $10^{-6}$, and its phase response is the one quoted; direct minimization agrees within its scatter of $2\times10^{-4}$.  Grids: $96^2$ and $144^2$ for $d=0.1$ ($192^2$ also at $E_g=80|U|$), $48^2$ and $96^2$ for $d=0.25$ and $0.5$; only agreeing digits are shown.  Numerical (\texttt{theta\_bdg\_scan.py}).}
\label{tab:thetabdg}
\centering
\begin{tabular}{cccccc}
\toprule
$d$ & $E_g/|U|$ & $1-{}$relaxed & frozen & $1+w_{12}d^2/\pi$ & $\frac12\dd\theta/\dd q_y$\\
\midrule
0.1 & 40 & $8\times10^{-5}$ & 1.03007 & 1.030153 & $-0.10000$\\
0.1 & 80 & $2\times10^{-5}$ & 1.03013 & 1.030153 & $-0.10000$\\
0.1 & 160 & $<10^{-5}$ & 1.03015 & 1.030153 & $-0.10000$\\
0.25 & 40 & $8\times10^{-5}$ & 1.17396 & 1.174056 & $-0.25000$\\
0.25 & 80 & $2\times10^{-5}$ & 1.17403 & 1.174056 & $-0.25000$\\
0.25 & 160 & $<10^{-5}$ & 1.17405 & 1.174056 & $-0.25000$\\
0.5 & 40 & $8\times10^{-5}$ & 1.65583 & 1.655965 & $-0.50000$\\
0.5 & 80 & $2\times10^{-5}$ & 1.65593 & 1.655965 & $-0.50000$\\
0.5 & 160 & $<10^{-5}$ & 1.65596 & 1.655965 & $-0.50000$\\
\bottomrule
\end{tabular}
\end{table}

\paragraph{Beyond mean field.}  The exact pair Hessian of the theta band is $G_\star/4\pi^2$ (Table~\ref{tab:pairmass}), the one-pair curvature converges to $4|U|\gbar$ (Table~\ref{tab:twobody}), and exact diagonalization confirms the quadratic filling law (Table~\ref{tab:ed}).  If the nondegeneracy found on these tori persists in the thermodynamic limit, the exact many-body stiffness of the theta band therefore sits on the floor in both realizations.

\paragraph{Ideal but not uniform.}  In its natural embedding the band obeys $\tr g=|\Om_{xy}|$ pointwise to $3\times10^{-16}$, while its Berry curvature ranges from $0$ to $-0.5075$ about the mean $-1/2\pi$, with a relative rms fluctuation of $1.047$, and $g_{xx}=g_{yy}$ fluctuates by the same amount ($d=0.5$; $96^2$ and $192^2$ agree; \texttt{theta\_inhomogeneity.py}).  For pairing within a single Chern band a uniform Landau level has no stiffness \cite{ZM2025,MC2024,GAF2025}; in the time-reversed channel this strongly inhomogeneous band sits exactly on the floor.

\subsection*{S5.2 The $\pi$-flux checkerboard band}
The flat Chern band of Hofmann, Berg, and Chowdhury, Model I of Gao, Han, and Khalaf \cite{HBC2020,GHK2026}, has the periodic-gauge Bloch Hamiltonian of Ref.~\cite{GHK2026} [their Eq.~(B1)] with $t_1=1$ and $t_2=1/\sqrt2$, in lattice coordinates $k_i=\bm k\cdot\bm a_i$, $\bm a_{1,2}=(1,\pm1)$; its two orbitals sit on different sites, at fractional displacement $(1/2,1/2)$.  Table~\ref{tab:hbc} gives its geometry (\texttt{hbc\_model.py}, identical at $96^2$ and $192^2$).  The natural embedding is the physical one, so $\Sigma=0$ and the band lies on the boundary curve of Fig.~1 at $R=1.163$, $\TLVP/2\pi=0.403$.  The Brillouin-zone average of its minimal metric, $\gbar=0.08580503$, reproduces the pair-mass value $2g/N_{\rm band}=0.085805$ of Ref.~\cite{GHK2026} (in their notation) to all quoted digits and the extrapolated bootstrap stiffness $\Ds/[|U|\nu(1-\nu)]=0.085799$ to a relative $7\times10^{-5}$.  Exact diagonalization (Table~\ref{tab:ed}) and the one-pair curvature (Table~\ref{tab:twobody}) agree.

\begin{table}[h]
\caption{The $\pi$-flux band ($96^2$ and $192^2$ agree).  Numerical.}
\label{tab:hbc}
\centering
\begin{tabular}{lc}
\toprule
$C$; orbital weights & $-1$; $0.5/0.5$\\
natural displacement $\bm d_\star$ (fractional) & $(1/2,1/2)$, the physical position of the second orbital\\
$G_\star/\pi$ (lattice coordinates) & $1.07825782\,I$\\
$R=\det G_\star/\pi^2C^2$; $\Torb/2\pi|C|=\TLVP/2\pi|C|$ & $1.16263993$; $0.40328642$\\
one-site frame: $\det G/\pi^2C^2$; $T/2\pi|C|$ & $2.32527987$; $1.15120800$\\
$\gbar$ & $0.08580503$\\
\bottomrule
\end{tabular}
\end{table}

\subsection*{S5.3 The honeycomb model of Li and Wu}
The $N=2$ model of Ref.~\cite{LiWu2026} lives on the honeycomb lattice: the lowest Landau level with one flux quantum per triangular unit cell, sampled at the A and B sites with equal weights, gives an exactly flat, ideal $C=1$ band below a dispersive band, and its relative hoppings match a Wannier model of the two top moir\'e bands of twisted MoTe$_2$ at the magic angle for three continuum parameter sets.  \texttt{liwu\_honeycomb.py} builds the flat band from Jacobi theta functions on the triangular lattice in the one-site periodic gauge (periodicity of $P$ to $3\times10^{-15}$).  Inversion through the hexagon center exchanges A and B and sends $\bm k\to-\bm k$; it maps the band to itself, $|P(-\bm k)|=|\sigma_xP(\bm k)\sigma_x|$ to $2\times10^{-15}$, so the sublattice weights are equal.  At $96^2$ and $192^2$: $C=-1$ (in our orientation), weights $0.5/0.5$, natural displacement $(-1/3,-2/3)$ in lattice coordinates, which is the physical displacement of B relative to A, and $R=1.0000000000$.  In mean-field theory, and exactly if its paired ground state is nondegenerate (not checked for this band), Theorem~(ii) therefore puts the time-reversed pair on the floor with $\Sigma=0$.  With both orbitals on one site, the same band would sit at $\TLVPcl/2\pi=0.99628450$ with $R=1$, but that is not the physical realization.

\subsection*{S5.4 An exactly vortexable elliptic family}
\begin{figure}[h]
\centering
\includegraphics[width=0.95\textwidth]{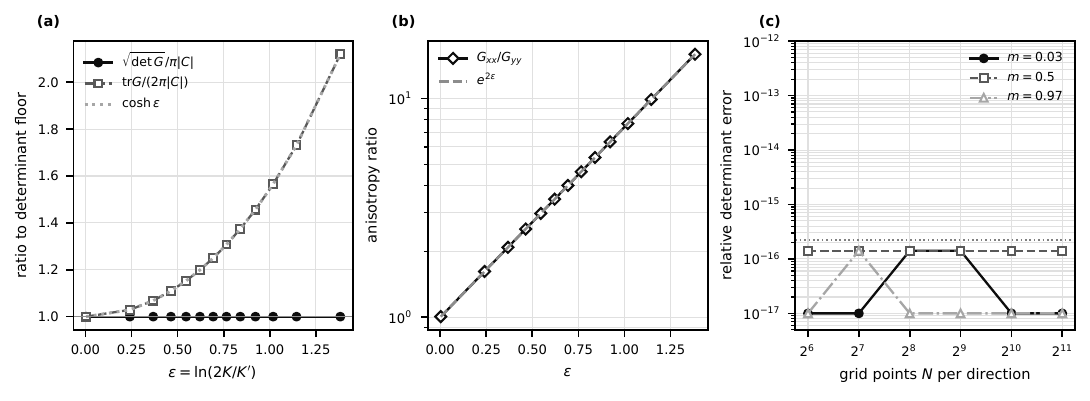}
\caption{Exactly vortexable $|C|=2$ elliptic family.  (a) $\sqrt{\det G}/\pi|C|$ stays at one while $\tr G/2\pi|C|$ grows as $\cosh\epsilon$.  (b) $G_{xx}/G_{yy}=\e^{2\epsilon}$.  (c) Saturation against grid size for three values of $m$.}
\label{fig:elliptic}
\end{figure}

Let $K=K(m)$ and $K'=K(1-m)$, and define the two-orbital, one-site band $u=(1,f_m)/\sqrt{1+|f_m|^2}$ with
\begin{equation}
f_m(\bm k)=m^{1/4}\,\mathrm{sn}\Bigl(\frac{2K}{\pi}k_x+\ii\frac{K'}{\pi}k_y\Big|m\Bigr).
\end{equation}
The Brillouin zone spans one period cell of $\mathrm{sn}$, which has two simple poles there, so $|C|=2$.  Holomorphy in the anisotropic variable gives $g=\frac12|\Om_{xy}|\,\mathrm{diag}(a_m/b_m,b_m/a_m)$ with $a_m=2K/\pi$ and $b_m=K'/\pi$, hence
\begin{equation}
G=\pi|C|\,\mathrm{diag}(\e^{\epsilon},\e^{-\epsilon}),\qquad\sqrt{\det G}=\pi|C|,\qquad\epsilon=\ln\frac{2K}{K'} .
\end{equation}
The frame saturates $\det G\ge\pi^2C^2$, so $G_\star=G$, $\Sigma=0$, and $\Torb=\TLVPcl=\TLVP=0$.  The identity $\mathrm{sn}(z+\ii K'|m)=1/[\sqrt m\,\mathrm{sn}(z|m)]$ gives $f_m(k_x,k_y+\pi)=1/f_m(k_x,k_y)$, which exchanges the two orbital weights, so uniform pairing holds for every $m$.  The determinant stiffness is pinned to the floor while the anisotropy $G_{xx}/G_{yy}=\e^{2\epsilon}$ changes exponentially (Fig.~\ref{fig:elliptic}): vortexability pins the determinant, not the shape.  Numerically (\texttt{make\_figS1\_elliptic.py}, $512^2$, 13 values of $m$): $|C|=2$, orbital-weight error $6\times10^{-17}$, floor residual $4\times10^{-16}$; the residual stays at roundoff from $64^2$ to $2048^2$, and \texttt{theta\_check.py} checks the Jacobi functions against a theta-function implementation to $10^{-14}$.

\subsection*{S5.5 Landau levels and mixtures}
For the $n$th Landau level with one flux quantum per magnetic unit cell, $C=1$ and the quantum metric is uniform, $g=\frac{2n+1}2\ell_B^2I$ with $2\pi\ell_B^2=1$, so $G=\pi(2n+1)I$; the corresponding $2n+1$ enhancement of the flat-band stiffness is in Ref.~\cite{PT2015}.  That the native frame is natural is shown here.  A magnetic translation $T_{\bm a}$ commutes with the band projector $\mathcal P$ and shifts $\hat r_j$ by a constant, which $(1-\mathcal P)\cdot\mathcal P$ annihilates, so the real-space density of the continuum linear functional of Sec.~S7.1, $\ell_j(\bm r)=2\,\mathrm{Re}\langle\bm r|(1-\mathcal P)\hat r_j\mathcal P|\bm r\rangle$, is translation invariant, hence constant, and it vanishes because the linear functional annihilates uniform translations.  Thus $\bm\eta^\star=0$, $G_\star=G$, the scalar continuum has no splitting defect, and, with the mean-field continuum stiffness of Sec.~S7.4,
\begin{equation}
\frac{\det\Ds}{\det\Ds^{\rm floor}}=(2n+1)^2,\qquad\TLVP=\TLVPcl=\Torb=4\pi\sqrt{n(n+1)},
\end{equation}
which places the time-reversed $n=1$ level on the boundary curve of Fig.~1 at $(2\sqrt2,9)$; Landau levels saturate the trade-off inequality.  For Dirac Landau levels the integrated trace differs: the $n$th Dirac level ($n\ge1$) is an equal mixture of the nonrelativistic levels $n-1$ and $n$ in a common gauge, and a direct computation gives $\tr G=2\pi\cdot2n$ rather than $2\pi(2n+1)$; and for modulated generalized Landau levels only $\det G_\star\le\pi^2(2n+1)^2C^2$ follows in general, from their integrated trace \cite{GLL2025}.

Consider next a two-component band whose components are the $n=0$ and $n=2$ levels with weights $p_0=p_2=1/2$, in the ladder gauge $|2,\bm k\rangle\propto(a^\dagger)^2|0,\bm k\rangle$.  The ladder operator commutes with magnetic translations, so both components carry the same Berry connection; the metric is $\frac12(g_0+g_2)$, $G=3\pi I$, $C=1$, and a relative displacement $\bm d$ only adds $p_0p_2\bm d\bm d^T$ times the Brillouin-zone area.  The native frame is natural and $R=9$, the value of the pure $n=1$ level.  Under component-diagonal contact attraction with uniform pairing, the two therefore have the same mean-field stiffness at equal pairing amplitude, and the mixture, with weight $1/2$ per component, is half as stiff at equal attraction.  An attraction on the total density, natural for co-located internal states, leaves inter-component terms, and then the theorem does not apply.

\subsection*{S5.6 The QWZ band}
\paragraph{Geometry.}  In the standard basis, $h_{\rm std}(\bm k)=\sin k_x\sigma_x+\sin k_y\sigma_y+(M-\cos k_x-\cos k_y)\sigma_z$, the lower band has $|C|=1$ for $-2<M<0$, but its orbital weights are unequal ($0.878/0.122$ at $M=-1.8$, $0.663/0.337$ at $M=-0.6$), so uniform pairing fails.  The Letter uses the rotated basis of Eq.~(8), obtained by conjugating $h_{\rm std}$ with the Hadamard matrix $(\sigma_x+\sigma_z)/\sqrt2$.  Three symmetries matter.  (a) $\sigma_xh(\bm k)\sigma_x=h(-\bm k)$ gives equal integrated weights for every $M$, so uniform pairing holds.  (b) $h(k_x,-k_y)=h(\bm k)^*$ and $h(-k_x,k_y)=\sigma_xh(\bm k)^*\sigma_x$ map the relative displacement $(d_x,d_y)$ to $(-d_x,d_y)$, so uniqueness gives $\bm d_\star=(0,d_\star)$.  (c) With both orbitals on one site the geometry is that of the standard basis and $C_4$ gives $G=aI$; but $C_4$ mixes the orbitals of the rotated basis, so it does not force $\bm d_\star=0$, and $G_\star=\mathrm{diag}(a,a-w_{12}d_\star^2)$.  In the standard basis, where $C_4$ acts by the orbital phases $\e^{-\ii\pi\sigma_z/4}$, the natural embedding is on site, as the symmetry theorem of Ref.~\cite{WRS2026} requires; at $M=-1$, $\det G_\star/\pi^2=2.030978$, the one-site value of the rotated basis, with weights $0.7456/0.2544$ (\texttt{qwz\_standard\_basis.py}, $96^2$ and $192^2$).  The two bases are different physical models, since the on-site attraction is diagonal in different orbitals.  Table~\ref{tab:qwzgeom} lists the geometry of the rotated basis.

\begin{table}[h]
\caption{The QWZ band in the pairing-symmetric basis (\texttt{qwzx\_geom.py}, analytic derivatives; $200^2$ and $400^2$ agree to the digits shown, and \texttt{check\_qwzx\_rank2.py} reproduces three rows independently).  $d_\star$ in lattice constants.  ``One site'' is $\det G/\pi^2=1+(\TLVP/2\pi)^2$ of the one-site realization, ``natural'' is $\det G_\star/\pi^2=1+(\Torb/2\pi)^2$.  Last column: weights in the standard basis.  Numerical.}
\label{tab:qwzgeom}
\centering
\begin{tabular}{rcccccccc}
\toprule
$M$ & $w_{12}$ & $d_\star$ & one site & natural & $\TLVP/2\pi$ & $\Torb/2\pi$ & $G_\star^{yy}/G_\star^{xx}$ & standard weights\\
\midrule
$-1.8$ & 8.1705 & 0.24070 & 1.43145 & 1.25117 & 0.6568 & 0.5012 & 0.8741 & 0.878/0.122\\
$-1.7$ & 8.0762 & 0.27759 & 1.40688 & 1.17192 & 0.6379 & 0.4146 & 0.8330 & 0.863/0.137\\
$-1.6$ & 7.9860 & 0.31136 & 1.43342 & 1.13837 & 0.6583 & 0.3720 & 0.7942 & 0.848/0.152\\
$-1.5$ & 7.8988 & 0.34317 & 1.48817 & 1.12696 & 0.6987 & 0.3563 & 0.7573 & 0.832/0.168\\
$-1.4$ & 7.8135 & 0.37367 & 1.56289 & 1.12875 & 0.7503 & 0.3588 & 0.7222 & 0.816/0.184\\
$-1.2$ & 7.6467 & 0.43221 & 1.76229 & 1.15869 & 0.8731 & 0.3984 & 0.6575 & 0.782/0.218\\
$-1.0$ & 7.4819 & 0.48896 & 2.03098 & 1.21952 & 1.0154 & 0.4685 & 0.6005 & 0.746/0.254\\
$-0.8$ & 7.3169 & 0.54493 & 2.39108 & 1.32166 & 1.1794 & 0.5671 & 0.5527 & 0.706/0.294\\
$-0.6$ & 7.1505 & 0.60040 & 2.89532 & 1.49923 & 1.3767 & 0.7066 & 0.5178 & 0.663/0.337\\
\bottomrule
\end{tabular}
\end{table}

Two physical realizations share this Bloch Hamiltonian.  With the orbitals on distinct sites, $\Vcal=\Ecal$ and $\TLVPcl=\Torb$; the published $\TLVP$ equals this value when the natural site configuration is collision free, as it is here.  With both orbitals on one site, the site-rigid optimum is attained trivially, $\TLVPcl=\TLVP$ is the one-site value, and $\Sigma=w_{12}\bm d_\star\bm d_\star^T$.  The mean-field problem is the same for both.

\paragraph{BdG.}  \texttt{bdg\_stiffness.py} uses the flattened band $h_\uparrow(\bm k)=E_g[1-P(\bm k)]$, its partner $h_\downarrow(\bm k)=h_\uparrow(-\bm k)^*$, and the attraction $-|U|\sum n_{\bm R\alpha\uparrow}n_{\bm R\alpha\downarrow}$, with pairing field $\Delta_{\bm R\alpha}=\Delta_\alpha\e^{2\ii\bm q\cdot\bm R}$ in the periodic gauge and $\hat\Delta=\mathrm{diag}(|\Delta_1|,|\Delta_2|\e^{\ii\theta})$.  The zero-temperature grand potential per cell is $\mathcal F=N_{\rm c}^{-1}\sum_{\bm k}\sum_n\min[E_n(\bm k;\bm q),0]+(|\Delta_1|^2+|\Delta_2|^2)/|U|$; Hartree shifts are orbital independent by symmetry (a) and absorbed in $\mu=0$, which is half filling up to corrections of order $|U|/E_g$.  Three Hessians are computed: \emph{relaxed}, minimizing $\mathcal F$ over $(|\Delta_1|,|\Delta_2|,\theta)$ at twists $\pm h$, $\pm2h$ (step $h=0.005$; the default $0.02$ of the script leaves an error of $1.2\times10^{-5}$ at $M=-1.8$, while $h=0.005$ and $0.0025$ agree to $10^{-7}$) in three directions with Richardson extrapolation; \emph{frozen}, the same differences with the order parameters held at their untwisted values, which is $A_e$ in the one-site frame; and \emph{Schur}, Eq.~\eqref{S:schur} from central differences at the untwisted saddle, with no minimization at finite twist.  Table~\ref{tab:qwzbdg} shows the results.  The relaxed stiffness follows the theorem with $\Torb$ to $3\times10^{-5}$ and not the one-site value, from which it differs by up to a factor $1.93$; the frozen Hessian is $\kappa_HG$ of the one-site frame; the Schur complement agrees with direct minimization to $6\times10^{-7}$; the phase response obeys Eq.~\eqref{S:phase}; and the stiffness is anisotropic although the band is $C_4$ symmetric, because the orbital-diagonal attraction breaks $C_4$ in this basis.  The residual is a finite-gap correction of order $(|U|/E_g)^2$ (Table~\ref{tab:qwzconv}).  The whole family is the same at $96^2$ and $144^2$, and at $48^2$ to within $3\times10^{-7}$ except at $M=-1.8$, where the band varies fastest in $\bm k$ and $48^2$ is too coarse ($1.4\times10^{-4}$).

\begin{table}[h]
\caption{Self-consistent BdG on the QWZ band, $E_g=80|U|$, half filling, $|\Delta|/|U|=0.25078$, twist step $0.005$; $96^2$ and $144^2$ agree to the digits shown.  Ratios are $\det(\cdot)/(\kappa_H\pi)^2$; ``theorem'' and ``one site'' from Table~\ref{tab:qwzgeom}.  The Schur evaluation agrees with the relaxed stiffness to $6\times10^{-7}$, and its phase response is the one quoted.  Numerical (\texttt{bdg\_stiffness.py}, \texttt{make\_fig3\_qwz.py}).}
\label{tab:qwzbdg}
\centering
\begin{tabular}{rcccccccc}
\toprule
$M$ & theorem & relaxed & one site & frozen & $G_\star^{yy}/G_\star^{xx}$ & $F^{yy}/F^{xx}$ & $d_\star$ & $\frac12\dd\theta/\dd q_y$\\
\midrule
$-1.8$ & 1.25117 & 1.25114 & 1.43145 & 1.43142 & 0.8741 & 0.8741 & 0.24070 & 0.24070\\
$-1.7$ & 1.17192 & 1.17190 & 1.40688 & 1.40685 & 0.8330 & 0.8330 & 0.27759 & 0.27759\\
$-1.6$ & 1.13837 & 1.13835 & 1.43342 & 1.43340 & 0.7942 & 0.7942 & 0.31136 & 0.31136\\
$-1.5$ & 1.12696 & 1.12694 & 1.48817 & 1.48814 & 0.7573 & 0.7573 & 0.34317 & 0.34317\\
$-1.4$ & 1.12875 & 1.12872 & 1.56289 & 1.56286 & 0.7222 & 0.7222 & 0.37367 & 0.37367\\
$-1.2$ & 1.15869 & 1.15867 & 1.76229 & 1.76225 & 0.6575 & 0.6575 & 0.43221 & 0.43221\\
$-1.0$ & 1.21952 & 1.21950 & 2.03098 & 2.03094 & 0.6005 & 0.6005 & 0.48896 & 0.48896\\
$-0.8$ & 1.32166 & 1.32163 & 2.39108 & 2.39103 & 0.5527 & 0.5527 & 0.54493 & 0.54493\\
$-0.6$ & 1.49923 & 1.49920 & 2.89532 & 2.89526 & 0.5178 & 0.5178 & 0.60040 & 0.60040\\
\bottomrule
\end{tabular}
\end{table}

\begin{table}[h]
\caption{Convergence of the relaxed determinant ratio at $M=-1$ toward the theorem value $1.219520$, twist step $0.005$.  The grid dependence, checked at $E_g=80|U|$, is $1\times10^{-6}$ between $24^2$ and $96^2$ and below $10^{-7}$ between $48^2$ and $96^2$.  Numerical (\texttt{bdg\_stiffness.py}, \texttt{run\_convergence}).}
\label{tab:qwzconv}
\centering
\begin{tabular}{rcccc}
\toprule
grid & $E_g/|U|$ & $|\Delta|/|U|$ & deviation ($10^{-5}$) & deviation$\times(E_g/|U|)^2$\\
\midrule
$48^2$ & 10 & 0.256408 & $-160.2$ & $-0.160$\\
$48^2$ & 20 & 0.253164 & $-39.07$ & $-0.156$\\
$48^2$ & 40 & 0.251572 & $-9.64$ & $-0.154$\\
$48^2$ & 80 & 0.250784 & $-2.39$ & $-0.153$\\
$48^2$ & 160 & 0.250391 & $-0.57$ & $-0.15$\\
$24^2$ & 80 & 0.250784 & $-2.28$ & $-0.146$\\
$96^2$ & 80 & 0.250784 & $-2.39$ & $-0.153$\\
\bottomrule
\end{tabular}
\end{table}

\section*{S6. Heterostrained chiral twisted bilayer graphene}

\subsection*{S6.1 Model}
We use LVP's chiral model with relaxation strain [their Eq.~(C8), with vanishing AA tunneling] \cite{LVP2023},
\begin{equation}
D=\begin{pmatrix}2\chi_+^\mu(-\ii\partial_\mu+A_{+\mu}) & \alpha_{\rm w}\mathcal T(\bm r)\\ \alpha_{\rm w}\mathcal T(-\bm r) & 2\chi_-^\mu(-\ii\partial_\mu+A_{-\mu})\end{pmatrix},\qquad
\mathcal T(\bm r)=\e^{-\ii\bm q_1\cdot\bm r}+\e^{2\pi\ii/3}\e^{-\ii\bm q_2\cdot\bm r}+\e^{-2\pi\ii/3}\e^{-\ii\bm q_3\cdot\bm r},
\end{equation}
with $\bm q_1=k_\theta(0,-1)$, $\bm q_{2,3}=k_\theta(\pm\sqrt3/2,1/2)$, $\alpha_{\rm w}=w_{\rm AB}/vk_\theta$ with $w_{\rm AB}$ the AB tunneling and $v$ the Dirac velocity, lengths in units of $k_\theta^{-1}$, and energies in units of $vk_\theta$; LVP write $\alpha$ and $U_1(\bm r)$, and Ref.~\cite{TKV2019} writes $\alpha$ and $U(\bm r)$, for $\alpha_{\rm w}$ and $\mathcal T(\bm r)$.  These are the interlayer phases of Ref.~\cite{TKV2019} and of LVP's Eq.~(16).  LVP's Eq.~(C4) defines $U_1$ with the opposite phases, which is consistent with their Hamiltonian, Eq.~(C3), whose lower-left tunneling block is the complex conjugate $\overline{U_1(-\bm r)}$.  Their Eq.~(C8), however, takes $U_1$ over without this conjugation.  With the phases of Eq.~(C4) inserted into Eq.~(C8), the unstrained model has no protected Dirac point and no magic angle for $0.3\le\alpha_{\rm w}\le2.5$, and with either relaxation sign there is no flat Chern band (\texttt{tbg\_phase\_convention.py}).  In Eq.~(C8) we therefore use the phases written above.  The flat band is the A-sublattice band, the lowest right singular vector of $D(\bm k)$.

The layers carry equal and opposite displacements $\bm u_\pm=\pm\bm u/2$.  This is the heterostrain of LVP's main text, the strain that ``the same relaxation mechanisms that led to suppressed AA tunneling'' produce, equal and opposite in the two layers; it leaves the moir\'e cell unchanged.  Uniform heterostrain imposed by a substrate distorts the moir\'e cell and is a different model, not computed here.  We keep only the first harmonic of the relative displacement, a truncation of ours,
\begin{equation}
\bm u(\bm r)=U_0\sum_{j=1}^3\sin(\bm K_j\cdot\bm r)\,\hat{\bm z}\times\hat{\bm K}_j ,
\end{equation}
where $\bm K_{1,2,3}$ are first-shell moir\'e reciprocal vectors related by $C_3$.  The field is $C_3$ symmetric, divergence free, and averages to zero.  With $u^\pm_{\mu\nu}=\frac12(\partial_\mu u^\pm_\nu+\partial_\nu u^\pm_\mu)$ the tetrad and the gauge field are
\begin{equation}
2\chi_\pm^\mu=(\delta^\mu_x+u_\pm^{\mu x})+\ii(\delta^\mu_y+u_\pm^{\mu y}),\qquad
\bm A_\pm=c_A\,(u^\pm_{xx}-u^\pm_{yy},\,-2u^\pm_{xy}),\qquad c_A=\frac{\sqrt3\,\beta_{\rm G}}{2ak_\theta},
\end{equation}
with $\beta_{\rm G}=3.14$ the logarithmic derivative of the nearest-neighbor hopping with respect to the bond length and $a=0.246$~nm.  The coefficient $\sqrt3\beta_{\rm G}/2a=\beta_{\rm G}/2a_{\rm cc}$ is the standard one \cite{VKG2010,NK2017}; LVP's Eq.~(C6) writes $\beta_{\rm G}/2a$ with $a$ the graphene lattice constant, which we run as a variant.  All strain-dependent quantities are evaluated at the twist angle $\theta_{\rm tw}=1.05^\circ$ ($k_\theta=0.3120$~nm$^{-1}$, $c_A=35.43$), and $\alpha_{\rm w}$ is then tuned to the magic value, as in LVP's Fig.~2.  At fixed $w_{\rm AB}$ the main-point value $\alpha_{\rm w}=0.8756$ would correspond to $\theta_{\rm tw}\approx0.70^\circ$, so the strain fields belong to $1.05^\circ$ while the tunneling is retuned.  We keep this parametrization from LVP.  $U_0=u_1ak_\theta$, with $u_1$ the first-harmonic amplitude in units of $a$.  At $u_1=0.12$ ($0.030$~nm) the strain of each layer, which is traceless because the field is divergence free, has principal values up to $\pm0.80\%$, that is, $[(u_{xx}-u_{yy})^2+4u_{xy}^2]^{1/2}$ reaches $1.60\%$, and the pseudomagnetic field $(\hbar/e)\nabla\times\bm A$ peaks at $81$~T, matching the ``peaks of $\sim80$~T'' that LVP quote from ab initio relaxation \cite{LVP2023,NK2017}; this is our main point.  The relaxation sign lowers the magic angle ($\alpha_{\rm w}$ rises from $0.5857$ to $0.8756$).  Plane waves with $|\bm K+\bm q_1/2|\le n_{\rm pw}\sqrt3k_\theta$ per layer, with $\bm K$ a moir\'e reciprocal vector, $n_{\rm pw}=5$ unless stated ($4$, $6$, $7$ as checks), and $18^2$ or $24^2$ momenta.  At the main point the flat band ($C=-1$) is separated from the remote bands by $0.291\,vk_\theta$.

\subsection*{S6.2 The Gram problem}
A displacement field $\bm\eta(\bm r,l)$, $l=\pm$, acts as $u_{\bm k}(\bm r,l)\to\e^{-\ii\bm k\cdot\bm\eta(\bm r,l)}u_{\bm k}(\bm r,l)$.  This is the continuum form of the Gram identity (Sec.~S7) with the layer as the orbital index of one site.  Minimizing over layer-resolved fields gives $G_\Ecal$ with $\det G_\Ecal=\pi^2C^2+\Torb^2/4$; minimizing over common fields gives $G_\Vcal$ with $\det G_\Vcal=\pi^2C^2+(\TLVPcl)^2/4$, because LVP's coordinates act on both layers at one point and their lifted displacements are dense among common periodic fields (Sec.~S7.3).  $\Sigma=G_\Vcal-G_\Ecal$, and $C_3$ symmetry makes all three tensors multiples of the identity.  We report $r_X-1=\det G_X/\pi^2C^2-1=(T_X/2\pi)^2$, a ratio of band geometry and not of stiffnesses here.  Two minimizations are used.  \emph{Fourier fields}, spanned by $1$, $\cos\bm Q\cdot\bm r$, $\sin\bm Q\cdot\bm r$ with $|\bm Q|\le n\sqrt3k_\theta$ (``fields to shell $n$''), separately per layer or common, with exact quadrature: each result is the exact minimum over a subspace and bounds the infimum from above in the Loewner order.  \emph{One free displacement per grid point}: the covariance form has no derivatives of $\bm\eta$, so the minimizer solves the second-kind equation $\bar w\,\bm\eta-\int_\BZ w_{\bm k}\langle w_{\bm k},\bm\eta\rangle\,\dd^2k=-\bm\beta$, with $w_{\bm k}=|u_{\bm k}|^2$, which on an $N_{\rm g}\times N_{\rm g}$ grid per layer is a diagonal matrix minus one of rank equal to the number of momenta, solved exactly by the Woodbury identity.  This Nystr\"om discretization is not variational; it converges as $N_{\rm g}$ grows and resolves the narrow structure near the minima of the layer densities.  Checks: for the unstrained magic band $\tr g=|\Om|$ pointwise to $2\times10^{-15}$, and $r-1$ and $\Sigma$ vanish to $10^{-14}$ in every frame, so that $\Torb/2\pi$ and $\TLVP/2\pi$ are below $10^{-7}$; with the gauge field alone $r-1$ and $\Sigma$ vanish to $3\times10^{-12}$ (Fourier) and $10^{-15}$ (grid), so that $T/2\pi$ is below $2\times10^{-6}$; the common-field minimum of the grid method agrees with the Fourier value to seven digits at every $N_{\rm g}$.

\subsection*{S6.3 Results}

\begin{table}[h]
\caption{Heterostrained chiral TBG at $u_1=0.12$ (peak field $81$~T).  ``Both methods'': Fourier fields and the grid solution.  Fourier entries for $\Torb$ are variational upper bounds, rounded up; the bound on $\Sigma$ is the converged $G_\Vcal$ minus the Fourier bound on $G_\Ecal$, rounded down; grid entries are estimates.  Numerical (\texttt{tbg\_sigma\_production.py}, \texttt{tbg\_grid\_minimizer.py}, \texttt{tbg\_diagnostics.py}, \texttt{tbg\_layer\_rank.py}, \texttt{tbg\_flatness.py}).}
\label{tab:tbgmain}
\centering
\begin{tabular}{lll}
\toprule
quantity & value & resolution\\
\midrule
magic $\alpha_{\rm w}$; residual bandwidth & $0.87564$; $1.1\times10^{-6}\,vk_\theta$ & $n_{\rm pw}=5$, $6$; $6^2$ to $12^2$ momenta\\
$C$; gap to remote bands & $-1$; $0.291\,vk_\theta$ & $18^2$, $24^2$ momenta\\
$r-1$, physical embedding & $7.2328\times10^{-6}$ & $n_{\rm pw}=5$ to $7$\\
$r_\Vcal-1$ & $6.4522\times10^{-6}$ & both methods, all grids\\
$\TLVP/2\pi$ & $2.540123\times10^{-3}$ & $n_{\rm pw}=5$, $6$, $7$; both methods\\
$\Torb/2\pi$, Fourier fields to shell 14 & $\le1.72\times10^{-4}$ & $1.7109$ ($n_{\rm pw}=5$), $1.7100$ ($6$)\\
$\Torb/2\pi$, grid $N_{\rm g}=480$ & $2.46\times10^{-5}$ & $n_{\rm pw}=6$, $7$\\
$\Torb/2\pi$, extrapolated & $2.3\times10^{-5}$ to $2.4\times10^{-5}$ & see text\\
$\Sigma/\pi$ & $\succeq3.21\times10^{-6}I$; grid $3.2258\times10^{-6}I$ & $n_{\rm pw}=5$, $6$\\
LVP's operator used as a field, $r-1$ & $1.5856\times10^{-5}$ & Sec.~S6.4\\
smallest $\det(I+\nabla\bm\eta_\Vcal)$ & $0.9988$ & $96^2$, $144^2$\\
layer density, min/mean; max/mean & $4.68\times10^{-6}$; $4.34$ & $96^2$, $144^2$; $96^2$, $128^2$\\
\bottomrule
\end{tabular}
\end{table}

\begin{figure}[h]
\centering
\includegraphics[width=0.95\textwidth]{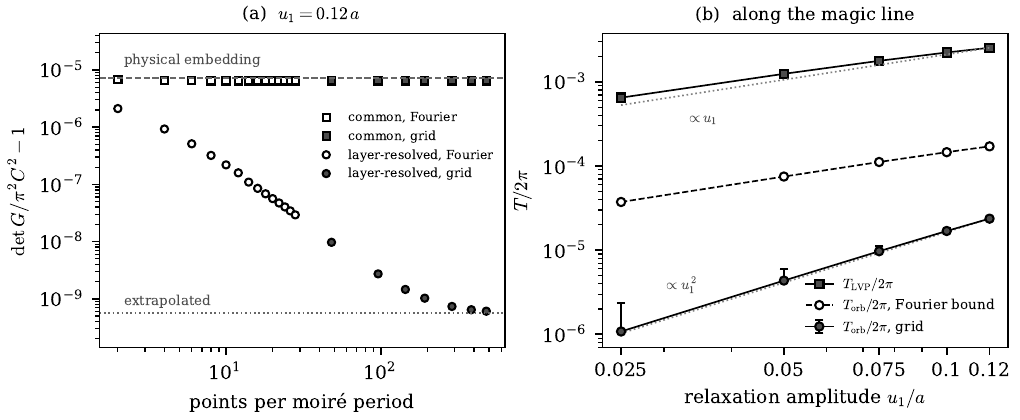}
\caption{Heterostrained chiral TBG.  (a) Main point: $\det G/\pi^2C^2-1$ against resolution in points per moir\'e period (Fourier fields to shell $n$ plotted at $2n$).  Squares: common fields; circles: layer-resolved fields; open symbols: Fourier fields ($n_{\rm pw}=5$, variational); filled symbols: one free displacement per grid point ($n_{\rm pw}=6$).  Dashed: physical embedding; dotted: extrapolation of the grid values.  (b) Along the magic line: $\TLVP/2\pi$ (filled squares), the bound on $\Torb/2\pi$ (open circles), and the grid estimate (filled circles; bars reach the values at $N_{\rm g}=480$).  Dotted guides $\propto u_1$ and $\propto u_1^2$.}
\label{fig:tbg}
\end{figure}

Common displacements lower $r-1$ from $7.233\times10^{-6}$ to $6.452\times10^{-6}$; the optimum is smooth and converged by shell 8.  Layer-resolved Fourier fields lower it below $2.93\times10^{-8}$ at shell 14, still falling roughly as $n^{-2}$.  The grid solution continues and levels off: $r_\Ecal-1=9.70$, $2.72$, $1.45$, $1.02$, $0.733$, $0.643$, and $0.607\times10^{-9}$ for $N_{\rm g}=48$, $96$, $144$, $192$, $288$, $384$, and $480$ ($n_{\rm pw}=6$), with a local exponent that falls from about $1.8$ to $0.26$.  A fit $r(N_{\rm g})=r_\infty+c_1N_{\rm g}^{-c_2}$ through the last three grids gives $\Torb/2\pi=2.367\times10^{-5}$; $n_{\rm pw}=7$ gives $2.366\times10^{-5}$ and $n_{\rm pw}=5$ gives $2.435\times10^{-5}$.  With $24^2$ momenta the grids $N_{\rm g}=192$, $288$, $384$ reproduce the $18^2$ values to all printed digits, and the fit through these three grids gives $2.339\times10^{-5}$: the spread comes from the fit window and the plane-wave cutoff, not from the momentum grid.  We quote $\Torb/2\pi\approx2.3$--$2.4\times10^{-5}\approx0.009\,\TLVP/2\pi$ as an estimate and $1.72\times10^{-4}$ as the bound (Fig.~\ref{fig:tbg}).  Along the magic line (Table~\ref{tab:tbgscan}) $\TLVP$ grows sublinearly in $u_1$ (local exponent from $0.94$ down to $0.70$), and the grid estimate follows $\Torb/2\pi\approx1.7\times10^{-3}u_1^2$ to within $5\%$; $\Sigma\neq0$ and $\Torb\le0.068\,\TLVP$ throughout.  The variants of Table~\ref{tab:tbgvar} change the model where LVP's choices could be questioned; in every case $\Sigma\neq0$ and $\Torb\le0.07\,\TLVP$, and the gauge field alone gives zero.

\begin{table}[h]
\caption{Along the magic line.  Bound: Fourier fields to shell 14 ($n_{\rm pw}=5$; $18^2$ and $24^2$ momenta with $96^2$ and $128^2$ real-space points agree to the digits shown), rounded up; $\Sigma$: converged $G_\Vcal$ minus that bound, rounded down.  Grid: extrapolated from $N_{\rm g}=288$, $384$, $480$ ($n_{\rm pw}=6$), one fit window at one resolution, so the second figure is uncertain by about one unit (at $u_1=0.12$ the spread over fit windows and cutoffs is shown); the entry for $u_1=0.025$, in parentheses, rests on a long extrapolation.  $B_{\rm peak}$: peak pseudomagnetic field ($64^2$ and $128^2$ real-space points agree to the digits shown).  Last column: largest eigenvalue ratio of the layer matrix $\mathcal L(\bm r)$ ($12^2$ and $18^2$ momenta agree).  Numerical.}
\label{tab:tbgscan}
\centering
\begin{tabular}{cccccccc}
\toprule
$u_1$ & $B_{\rm peak}$ (T) & $\alpha_{\rm w}$ & $\TLVP/2\pi$ ($10^{-3}$) & $\Torb/2\pi$ bound ($10^{-5}$) & grid ($10^{-5}$) & $\Sigma/\pi\succeq$ ($10^{-6}$) & $\lambda_{\min}/\lambda_{\max}$ ($10^{-6}$)\\
\midrule
$0.025$ & $17$ & $0.63774$ & $0.64890$ & $3.74$ & $(0.11)$ & $0.209$ & $0.38$\\
$0.05$ & $34$ & $0.69369$ & $1.24480$ & $7.49$ & $0.43$ & $0.771$ & $1.9$\\
$0.075$ & $51$ & $0.75400$ & $1.77662$ & $11.2$ & $0.97$ & $1.57$ & $5.1$\\
$0.1$ & $68$ & $0.81928$ & $2.23402$ & $14.6$ & $1.7$ & $2.48$ & $11$\\
$0.12$ & $81$ & $0.87564$ & $2.54012$ & $17.2$ & $2.3$--$2.4$ & $3.21$ & $18$\\
\bottomrule
\end{tabular}
\end{table}

\begin{table}[h]
\caption{Variants at $u_1=0.12$, Fourier fields to shell 14, $n_{\rm pw}=5$ ($18^2$ and $24^2$ momenta agree); units and rounding as in Table~\ref{tab:tbgscan}, last column computed before rounding.  The velocity coefficient multiplies the strain in the tetrad: $1$ is LVP's geometric term, $1-\beta_{\rm G}/2$ adds the change of the hopping with bond length \cite{dJSV2012,dJMV2013}.  ``Zero'': $r-1$ and $\Sigma$ below $3\times10^{-12}$ in magnitude, so $T/2\pi$ below $2\times10^{-6}$.  Numerical.}
\label{tab:tbgvar}
\centering
\begin{tabular}{lccccc}
\toprule
variant & $\alpha_{\rm w}$ & $\TLVP/2\pi$ ($10^{-3}$) & $\Torb/2\pi\le$ ($10^{-5}$) & $\Sigma/\pi\succeq$ ($10^{-6}$) & $(\Torb/\TLVP)^2\le$ ($10^{-3}$)\\
\midrule
main point & $0.87564$ & $2.5401$ & $17.2$ & $3.21$ & $4.54$\\
velocity coefficient $1-\beta_{\rm G}/2$ & $0.87564$ & $1.4479$ & $9.73$ & $1.04$ & $4.51$\\
gauge coefficient $\beta_{\rm G}/2a$ & $0.73979$ & $2.8773$ & $18.0$ & $4.12$ & $3.89$\\
opposite relaxation sign & $0.38026$ & $3.4923$ & $15.1$ & $6.08$ & $1.87$\\
tetrad only & $0.58566$ & $3.2238$ & $17.8$ & $5.18$ & $3.05$\\
gauge field only & $0.87564$ & zero & zero & zero & \\
\bottomrule
\end{tabular}
\end{table}

\subsection*{S6.4 Structure}
\emph{The gauge field alone keeps the band ideal and the layer spinor fixed} (proposition and proof in the End Matter of the Letter; the vortexability part is LVP's).  Numerically, for the unstrained and gauge-only bands the eigenvalue ratio of the layer matrix $\mathcal L(\bm r)=\int_\BZ\psi_{\bm k}(\bm r)\psi_{\bm k}(\bm r)^\dagger\dd^2k$ stays below $2\times10^{-14}$ and $2.4\times10^{-11}$, and each layer density vanishes, to $2\times10^{-12}$ of its mean, at one of the two stacking points other than AA.

\emph{The tetrad rotates the layer spinor.}  With the tetrad, $\mathcal L(\bm r)$ has rank two: its eigenvalue ratio reaches $1.81\times10^{-5}$ at the main point and grows faster than $u_1^2$ (local exponents from $2.3$ to $2.8$ along the magic line), and the zeros of the layer densities are lifted to $4.68\times10^{-6}$ of the mean.  For a vortex function diagonal in the layer index, $\tilde z=\mathrm{diag}(z_+,z_-)$, the commutator with $D$ is the multiplication operator
\begin{equation}
\mathcal C(\bm r)=[D,\tilde z]=\begin{pmatrix}-2\ii\chi_+^\mu\partial_\mu z_+ & \alpha_{\rm w}\mathcal T(\bm r)(z_--z_+)\\ \alpha_{\rm w}\mathcal T(-\bm r)(z_+-z_-) & -2\ii\chi_-^\mu\partial_\mu z_-\end{pmatrix}.
\end{equation}
If the band were exactly the kernel of $D$, invariance would require $\mathcal C(\bm r)\psi_{\bm k}(\bm r)=0$ for every $\bm k$; where $\mathcal L(\bm r)$ has rank two, which is almost everywhere because zero modes of $D$ are real analytic, so that $\det\mathcal L$ is real analytic and, not being identically zero, vanishes only on a set of measure zero, this forces $\mathcal C(\bm r)=0$, hence $z_+=z_-\equiv z$ wherever $\mathcal T\neq0$ (and by continuity everywhere) and $\chi_+^\mu\partial_\mu z=\chi_-^\mu\partial_\mu z=0$.  At first order in strain the determinant of the $2\times2$ matrix with rows $\chi_+$ and $\chi_-$ is proportional to $(u_{xx}-u_{yy})+2\ii u_{xy}$ of one layer, so the two conditions are independent wherever the traceless strain is nonzero, as LVP note for a common vortex function \cite{LVP2023}.  For the first-harmonic field the traceless strain is a nonzero trigonometric polynomial and vanishes only on a set of measure zero, so $\nabla z=0$ almost everywhere and $z$ would be constant, which a vortex function, growing linearly across the lattice, cannot be.  The layer densities stay positive, so the Fredholm argument of Sec.~S7.4, in its form for a density bounded away from zero, shows that the layer-resolved minimum is attained, and $\Torb=0$ would produce such a $\tilde z$.  This makes $\Torb>0$ the expected outcome, which the grid solution finds; it is not a proof for a band that misses exact flatness by $1.1\times10^{-6}\,vk_\theta$.

\emph{LVP's layer-dependent operator.}  LVP write that the opposite strains prevent a single vortex function and that ``it is possible to define a generalized `layer dependent vortex operator' $\tilde z={\rm diag}(z_+,z_-)$ that satisfies $[D,\tilde z]=0$ and $\tilde z\psi=P\tilde z\psi$'' \cite{LVP2023}.  With the crystal coordinates of the two layers, $z_\pm=\bm c\cdot(\bm r-\bm u_\pm)$, the diagonal entries of $\mathcal C$ vanish at first order in strain, but the off-diagonal entries are $\pm\alpha_{\rm w}\mathcal T(\pm\bm r)\,\bm c\cdot(\bm u_+-\bm u_-)\neq0$.  Used as a displacement field, $\bm\eta_\pm=-\bm u_\pm$, it gives $r-1=1.5856\times10^{-5}$, more than the physical embedding's $7.2328\times10^{-6}$; the layer-resolved optimum, $r-1\approx6\times10^{-10}$, is a different field.

\emph{The common minimizer is an admissible coordinate.}  $\max|\bm\eta_\Vcal|=3.06\times10^{-3}k_\theta^{-1}$ ($0.010$~nm), $\max|\nabla\bm\eta_\Vcal|=0.016$, and $\min\det(I+\nabla\bm\eta_\Vcal)=0.9988$ (shells 8 and 14, $96^2$ and $144^2$ points), so the Jacobian criterion of Sec.~S7.5 holds and $\TLVPcl=\TLVP$.  The layer-resolved optimum instead concentrates next to the near-zeros of the layer densities: its largest displacement grows from $0.0141$ to $0.0498\,k_\theta^{-1}$ between shells 4 and 14 while its location approaches the density minimum, which is why the Fourier bound decreases slowly.

\emph{Residual bandwidth.}  LVP state that the position-dependent velocities do not change the band energies, so that an exactly flat band survives at a renormalized angle \cite{LVP2023}.  Minimizing $\max_{\bm k}\sigma_{\min}(D(\bm k))$ over $\alpha_{\rm w}$ gives about $5\times10^{-9}\,vk_\theta$ for the unstrained band and $2.5\times10^{-9}\,vk_\theta$ with the gauge field alone, but about $1.9\times10^{-6}\,vk_\theta$ with the tetrad alone ($1.84$ and $1.87\times10^{-6}$ on $9^2$ and $12^2$ momentum grids) and $1.1\times10^{-6}\,vk_\theta$ for the full model, the same for $n_{\rm pw}=5$ and $6$.  With the tetrad alone the bandwidth grows quadratically in $u_1$; for the full model it grows more slowly (local exponents $1.8$ and $1.5$ between $u_1=0.03$, $0.06$, and $0.12$).  The crystal-frame argument removes the tetrad of one layer at a time but not of both at the same point, where the tunneling acts.  With $\hbar v\approx0.66$~eV\,nm, $vk_\theta\approx0.2$~eV and the bandwidth is about $0.2~\mu$eV; it matters only for hypotheses that need exact flatness, and part (i) of the theorem needs only an isolated band.

\emph{Scope.}  The layer-resolved band density varies from $4.7\times10^{-6}$ to $4.3$ times its mean, so uniform pairing fails badly and part (ii) says nothing about the stiffness of twisted bilayer graphene.  The model is LVP's (chiral limit, tetrad and gauge field at first order in strain, tunneling not relaxed), with the interlayer phases of Ref.~\cite{TKV2019} and of their Eq.~(16), in place of those that their Eq.~(C8) takes over from Eq.~(C4), the gauge coefficient $\sqrt3\beta_{\rm G}/2a$ in place of their $\beta_{\rm G}/2a$ (which we run as a variant), and the relaxation field restricted by us to its first harmonic.

\section*{S7. The continuum}

\subsection*{S7.1 Gram identity without a range assumption}
Let $x=(\bm r,\sigma)$ denote a point of the unit cell with any internal index, $w_{\bm k}(x)=|u_{\bm k}(x)|^2$, and let a real periodic field act as $u_{\bm k}(x)\to\e^{-\ii\bm k\cdot\bm\eta(x)}u_{\bm k}(x)$.  With the covariance form
\begin{equation}
\mathfrak w(f,g)=\int_\BZ\dd^2k\Bigl[\int\dd x\,w_{\bm k}fg-\Bigl(\int\dd x\,w_{\bm k}f\Bigr)\Bigl(\int\dd x\,w_{\bm k}g\Bigr)\Bigr]
\end{equation}
and $\ell_j(f)=-\int_\BZ\dd^2k\,\mathrm{Im}\int\dd x\,u^*_{\bm k}v_jf$, $v_j=(1-P_{\bm k})\partial_ju_{\bm k}$, the expansion of Eq.~\eqref{S:gshift} gives $G_{ij}[\bm\eta]=G^0_{ij}+\ell_j(\eta_i)+\ell_i(\eta_j)+\mathfrak w(\eta_i,\eta_j)$.  Positivity of $G_{jj}[tf]$ for all real $t$ gives $|\ell_j(f)|^2\le G^0_{jj}\,\mathfrak w(f,f)$, so $\ell_j$ is continuous in the seminorm of $\mathfrak w$.  On the Hilbert completion $\mathfrak H$ of the fields modulo $\ker\mathfrak w$, with canonical map $J$, the Riesz representation $\ell_j(f)=\langle\mathfrak h_j,Jf\rangle$ gives
\begin{equation}
G_{ij}[\bm\eta]=G^{\inf}_{ij}+\langle J\eta_i+\mathfrak h_i,J\eta_j+\mathfrak h_j\rangle,\qquad G^{\inf}_{ij}=G^0_{ij}-\langle\mathfrak h_i,\mathfrak h_j\rangle,
\label{S:contgram}
\end{equation}
so $G[\bm\eta]\succeq G^{\inf}$, $G^{\inf}$ is the Loewner infimum (since $J$ has dense range), $\det G^{\inf}\ge\pi^2C^2$, and $\inf_{\bm\eta}\det G[\bm\eta]=\det G^{\inf}$.  If $\bar w(x)=\int_\BZ w_{\bm k}(x)\dd^2k$ is bounded, $\mathfrak w(f,f)\le\|\bar w\|_\infty\|f\|^2_{L^2}$ and the metric is $L^2$ continuous.  (Unlike $\bar w_\alpha$ of Sec.~S4.1, this density carries no factor $(2\pi)^{-2}$.)

\subsection*{S7.2 Internal components}
Let $\Vcal$ be the fields independent of the internal index and $P_\Vcal$ the orthogonal projector onto the closure of $J\Vcal$.  Minimizing Eq.~\eqref{S:contgram} over common fields gives $G_\Vcal=G^{\inf}+\Sigma_{\rm cont}$ with $(\Sigma_{\rm cont})_{ij}=\langle(1-P_\Vcal)\mathfrak h_i,(1-P_\Vcal)\mathfrak h_j\rangle\succeq0$, the counterpart of Eq.~\eqref{S:sigma}.  A scalar band has $\Sigma_{\rm cont}=0$.

\subsection*{S7.3 Periodic density lemma}
\emph{Lemma.}  For $1\le p<\infty$, the lifted displacements $\widetilde\phi-\mathrm{id}$ of orientation-preserving torus diffeomorphisms isotopic to the identity are dense in $L^p(\mathbb T^2;\mathbb R^2)$.  This is a torus version of the density proved by Brenier and Gangbo for bounded Euclidean domains \cite{Brenier2003}.

\emph{Proof.}  Bounded continuous fields are dense in $L^p$, so fix a continuous target $\bm\eta$ and $\epsilon>0$.  Choose a partition of the torus into cells of diameter $\delta$ with $|\bm\eta(\bm r)-\bm\eta(x_a)|<\epsilon$ in each cell, a center $x_a$ in every cell, pairwise disjoint small core disks $D_a$ about the $x_a$, and pairwise disjoint target disks $C_a$ of the same radius whose centers lie arbitrarily close to the projections of $x_a+\bm\eta(x_a)$, separating coincident targets by small perturbations.  Fix the target centers, and choose collision-free paths of the centers to the targets, with the lattice winding of each strand chosen so that its lift ends near $x_a+\bm\eta(x_a)$; windings can be inserted one strand at a time in the punctured torus.  Then take the common radius of the $D_a$ and $C_a$ below half the least distance between distinct strands at equal times.  Transporting the disks $D_a$ along these paths and adjusting them to $C_a$ gives, by isotopy extension, $\phi_2\in\mathrm{Diff}_0^+(\mathbb T^2)$, whose lift has a displacement bound $B_2$ fixed by the choices so far.  Now choose in each cell a closed topological disk $B_a\supset D_a$ whose complement in the cell is arbitrarily thin, and an orientation-preserving diffeomorphism $\phi_1$, supported in the cells, that maps each $B_a$ into $D_a$ and moves points by at most $\delta$.  Then the lift of $\phi=\phi_2\circ\phi_1$ obeys $|\widetilde\phi(\bm r)-\bm r-\bm\eta(\bm r)|\le B$ everywhere, with a constant $B$ that depends on $B_2$, $\delta$, and $\|\bm\eta\|_\infty$ but not on the thickness of the gaps.  On $\bigcup_aB_a$ the displacement is $\bm\eta(x_a)+O(\delta+\epsilon)=\bm\eta(\bm r)+O(\delta+\epsilon)$.  With $B$ fixed, make the gaps so thin that their total measure is below $(\epsilon/B)^p$; the $L^p$ norm of $\widetilde\phi-\mathrm{id}-\bm\eta$ over the gaps is then at most $\epsilon$.  Letting $\delta,\epsilon\to0$ proves the lemma. $\square$

For a scalar band the lemma and $L^2$ continuity give $\inf_{\phi\in\Dcal}\det G_\phi=\det G^{\inf}$: the relaxed LVP infimum equals the continuum natural infimum.  For several internal components the same argument gives $\det G_\Vcal$.

\subsection*{S7.4 Continuum stiffness}
For an isolated, exactly flat, time-reversed scalar band with contact attraction and a spatially uniform pair field, the flat-band gap equation is uniform only if $\bar w(\bm r)=\bar w>0$, which is part of the uniform-pairing hypothesis.  More generally, let $0<\inf\bar w\le\sup\bar w<\infty$, with $\bar w(x)=\int_\BZ w_{\bm k}(x)\dd^2k$.  The covariance operator is $W=\bar w^{1/2}(1-\widetilde K)\bar w^{1/2}$, with $(\widetilde Kf)(x)=\int\bar w(x)^{-1/2}K(x,x')\bar w(x')^{-1/2}f(x')\dd x'$ and $K(x,x')=\int_\BZ w_{\bm k}(x)w_{\bm k}(x')\dd^2k$; for uniform $\bar w$ this is $\bar w(1-\widetilde K)$.  For a smooth isolated band $\widetilde K$ is compact (Hilbert--Schmidt) and smoothing, the linear term is orthogonal to $\ker W$ by the inequality of Sec.~S7.1, and the Fredholm alternative for $1-\widetilde K$ gives a smooth solution of $W\eta^\star_j=-\beta_j$, with $\beta_j(x)=-\int_\BZ\mathrm{Im}[u^*_{\bm k}(x)v_j(x)]\,\dd^2k$ the density of $\ell_j$, unique modulo $\ker W$: the field minimum is attained.  Uniform $\bar w$ is needed only for the stiffness relation below; attainment holds for any density bounded away from zero, which is the case used for the layers of Sec.~S6.  Zeng and Millis give the contact-interaction flat-band stiffness for diagonal components \cite{ZM2025}; the local pair phase is $\theta_{\bm q}(\bm r)=2\bm q\cdot\bm\eta(\bm r)+o(|\bm q|)$, and applying their result to every twist direction and polarizing gives $\Ds(0)=\kappa_0G^{\inf}$.  With the lemma this yields Eq.~(7) of the Letter.  This continuum stiffness is the mean-field one; the argument of Sec.~S4.2 uses only $\bar n^2=\bar w\bar n$ and the $\eta$-pairing SU(2), which hold here as well, but we have not checked the continuum two-body limit, so Eq.~(7) and the Landau-level statements of Sec.~S5.5 are mean-field statements.

\subsection*{S7.5 Attainment and actual vortexability}
Put $F^\star(\bm r)=\bm r+\bm\eta^\star(\bm r)$.  If constants are the only covariance zero modes, all minimizers differ by translations and
\begin{equation}
\TLVPcl=\TLVP\quad\Longleftrightarrow\quad\det[I+\nabla\bm\eta^\star(\bm r)]>0\ \text{for all }\bm r .
\end{equation}
A positive Jacobian makes $F^\star$ an orientation-preserving local diffeomorphism of the torus of degree one, hence a global diffeomorphism, and compactness gives LVP's uniform nondegeneracy; conversely, an admissible minimizer has a positive Jacobian.  With further zero modes the criterion is the existence of one smooth minimizer with this property.  A sufficient condition is $\sup_{\bm r}\|\nabla\bm\eta^\star\|<1$.  The heterostrained bilayer of Sec.~S6 satisfies the criterion with margin.

\section*{S8. Finite temperature}

Under the hypotheses of part (ii) the projected Hamiltonian has the exact $\eta$-pairing SU(2) symmetry of Sec.~S4.2 \cite{Tovmasyan2016}.  Its paired state is a pseudospin ferromagnet whose Goldstone mode disperses quadratically \cite{HHBK2024,HKH2026}, and in two dimensions the Mermin--Wagner theorem \cite{MerminWagner1966} forbids order at any $T>0$: $T_c=0$ although $\Ds(0)>0$.  A finite $T_c$ needs a symmetry-breaking anisotropy, of energy scale $\varepsilon_a$, from residual bandwidth, interband processes, or further interactions, and its sign matters.  Easy-plane breaking favors superconductivity with a Berezinskii--Kosterlitz--Thouless transition; easy-axis breaking favors charge order or phase separation.  In quantum Monte Carlo on the $\pi$-flux band, interband processes and residual bandwidth favor superconductivity, whereas a nearest-neighbor attraction of $0.1|U|$ drives phase separation \cite{HBC2020}.

For a weak easy-plane anisotropy, Han \emph{et al.}\ estimate $k_BT_c\simeq4\pi|\Upsilon|/\ln(|\Upsilon|/\varepsilon_a)$, with $|\Upsilon|=\sqrt{\det\Upsilon}$ the geometric mean of the pseudospin stiffness $\Upsilon$, from the crossover of the anisotropic O(3) model to XY behavior \cite{HHBK2024,NP1977,HikamiTsuneto1980}.  With exactness, $\Upsilon$ is proportional to $\Ds(0)$ at fixed filling.  For two bands compared at equal pairing amplitude, filling, and $\varepsilon_a$,
\begin{equation}
\frac{T_c}{T_c^{\rm floor}}=\sqrt R\,\frac{\ln(|\Upsilon_{\rm floor}|/\varepsilon_a)}{\ln(\sqrt R|\Upsilon_{\rm floor}|/\varepsilon_a)}=\sqrt R\Bigl[1-\frac{\ln\sqrt R}{\ln(|\Upsilon_{\rm floor}|/\varepsilon_a)}+\dots\Bigr],\qquad
\Bigl(\frac{T_c}{T_c^{\rm floor}}\Bigr)^2=R\,\bigl[1+O\bigl(1/\ln(|\Upsilon|/\varepsilon_a)\bigr)\bigr].
\end{equation}
This assumes that $\varepsilon_a$ is an energy scale that does not change with the band geometry; if instead the anisotropy is a fixed fraction of $|\Upsilon|$, the logarithms are equal and $T_c/T_c^{\rm floor}=\sqrt R$ at this order.  Because $T_c\to0$ as $\varepsilon_a\to0$, thermal depletion of the stiffness does not enter at this order.  The formula is that of Ref.~\cite{HHBK2024}; we claim only this corollary, and nothing for strong breaking.  Inserting the mean-field stiffness into the Nelson--Kosterlitz criterion \cite{NelsonKosterlitz1977} would give a crossing temperature proportional to $|U|n_\phi$, which is not a transition temperature in this regime.

\section*{S9. Robustness}

The BdG runs of this section use the twist step $0.02$ of \texttt{bdg\_stiffness.py}; repeating every $96^2$ run with the step $0.005$ (\texttt{step\_check.py}) reproduces all quoted digits.

\subsection*{S9.1 Dispersion: a weighted Gram theorem}
For an isolated band with dispersion $\varepsilon_{\bm k}$ the stiffness weight depends on $\bm k$.  Tam and Peotta note that natural positions exist also for a dispersive isolated band, with the weight $E_{\bm k}^{-1}\tanh(E_{\bm k}/2k_BT)$ inserted in their Brillouin-zone integrals \cite{TP2024}, and Prijon, Huber, and Huhtinen derive a Chern floor for the dispersive geometric stiffness \cite{Prijon2026}.  For a bounded weight $\kappa(\bm k)\ge0$, define $G^\kappa_e=\int\kappa g_e$, $W^\kappa=\int\kappa(\mathrm{diag}\,P-P\circ P^*)$ and $\bm b^\kappa=\int\kappa\bm\beta$.

\emph{Theorem (weighted Gram identity).}  $G^\kappa_e=G^\kappa_0+\bm\eta^T\bm b^\kappa+\bm b^{\kappa T}\bm\eta+\bm\eta^TW^\kappa\bm\eta$ with $W^\kappa\succeq0$; each column of $\bm b^\kappa$ lies in $\operatorname{Ran}W^\kappa$; and $G^\kappa_e=G^\kappa_\star+\Gamma^\kappa(\bm\eta-\bm\eta^\kappa_\star)\succeq G^\kappa_\star$.  If $\kappa>0$ almost everywhere, $\ker W^\kappa=\ker W^1$.  The proof is that of Sec.~S3.1 with the pointwise identity multiplied by $\kappa\ge0$.

\emph{Physical weight.}  In mean-field theory, in the isolated-band limit with uniform pairing, the relaxed stiffness is $F_{qq}=D_{\rm conv}+G^\kappa_\star$ with $\kappa_{\bm k}=\Delta^2\tanh(E_{\bm k}/2k_BT)/(\pi^2E_{\bm k})$ and the embedding-independent conventional term $D_{\rm conv}$ \cite{Liang2017,TP2024}; time reversal makes the amplitudes even and the phases odd in $\bm q$, so only the relative phases relax, and $\dd\theta_\alpha/\dd\bm q=2(\bm\eta^\kappa_{\star\alpha}-\bm\eta_\alpha)$.  For a flat band at $T=0$, $E_{\bm k}=|U|n_\phi/2$ at every filling, so $\kappa=\Delta^2/(\pi^2E)=\kappa_0$.

\emph{Two-sided bound.}  With $\kappa_{\min}\le\kappa\le\kappa_{\max}$, $\kappa_{\min}G_\star\preceq G^\kappa_\star\preceq G^\kappa_{e_\star}\preceq\kappa_{\max}G_\star$, so
\begin{equation}
\kappa_{\min}^2\bigl(\pi^2C^2+\tfrac14\Torb^2\bigr)\le\det G^\kappa_\star\le\kappa_{\max}^2\bigl(\pi^2C^2+\tfrac14\Torb^2\bigr)\le\kappa_{\max}^2\bigl(\pi^2C^2+\tfrac14(\TLVPcl)^2\bigr),
\end{equation}
and at $T=0$, $D_{\rm conv}\succeq0$ gives $\det F_{qq}\ge\det G^\kappa_\star$.  The lower half at $\Torb=0$ is the floor of Ref.~\cite{Prijon2026}.

\emph{Vortexability no longer pins the stiffness.}  If the flat-band natural frame is ideal, $g_{e_\star}=\frac12|\Om_{e_\star}|\omega$, then $G^\kappa_{e_\star}$ has the shape $\omega$, but $G^\kappa_\star=G^\kappa_{e_\star}-\bm b^{\kappa T}_{e_\star}W^{\kappa+}\bm b^\kappa_{e_\star}$ differs from it unless $\bm b^\kappa_{e_\star}=0$.  The weighted curvature integral is not embedding invariant: with $\langle\eta_j\rangle_{\bm k}=\sum_\alpha w_\alpha(\bm k)\eta_{\alpha j}$ the shift of the Berry connection, $\int\kappa\Om_e=\int\kappa\Om-\int(\partial_x\kappa\langle\eta_y\rangle_{\bm k}-\partial_y\kappa\langle\eta_x\rangle_{\bm k})$, so no exact analog of the LVP identity holds when $\nabla\kappa\neq0$, and only the two-sided bound survives.  The weighted metric evaluated at the flat-band natural embedding overestimates $G^\kappa_\star$ only at second order in the variation of $\kappa$.

\emph{Numerics.}  \texttt{bdg\_dispersive.py} replaces the flattened band by $h(\bm k)=\varepsilon_{\bm k}P(\bm k)+E_g[1-P(\bm k)]$, $\varepsilon_{\bm k}=t_b(\cos k_x+\cos k_y)$, $\mu=0$, $T=0$; both test bands have $\sigma_xP(\bm k)\sigma_x=P(-\bm k)$, so uniform pairing holds by symmetry.  Table~\ref{tab:disp} shows that the relaxed stiffness and the phase response follow the weighted prediction, with residuals that roughly halve when $E_g$ doubles, and that the phase response measures the weighted displacement $d^\kappa_\star$, not the flat-band $d_\star$.  For the vortexable theta band with dispersion ($E_g=320|U|$) the weighted metric in the frame where the band is holomorphic is isotropic, $0.119605\,I$ after adding $D_{\rm conv}$, while relaxation lowers the $yy$ component to $0.119106$ (BdG $0.119045$): vortexability no longer pins the stiffness.  \texttt{tdep.py} follows the natural displacement with temperature (QWZ, $M=-1$, $t_b=0.25$): it drifts from $0.43297$ at $T=0$ to $0.44104$ as $T\to T_c$ in the isolated-band theory, and the finite-temperature BdG reproduces the drift ($0.43322$, $0.43452$, $0.43979$ at $T/T_c=0$, $0.5$, $0.9$, against $0.43303$, $0.43432$, $0.43958$ at the BdG gap), with a constant offset of $2\times10^{-4}$ that is the finite-gap residual (BdG at $48^2$ and $96^2$, isolated-band theory at $192^2$ and $288^2$, agree).

\begin{table}[h]
\caption{Dispersive bands, $T=0$; $48^2$ and $96^2$ agree to the digits shown in every row.  $F_{yy}$: relaxed BdG (Schur evaluation) against $D_{\rm conv}+G^\kappa_\star$.  $(A_0-F)_{yy}$: frozen minus relaxed against $\Gamma^\kappa(\bm\eta^\kappa_\star)$.  Phase response (Schur) against $d^\kappa_\star$; $d_\star$ is the flat-band value.  Numerical (\texttt{bdg\_dispersive.py}).}
\label{tab:disp}
\centering
\begin{tabular}{lcccccccccc}
\toprule
band & $t_b$ & $E_{\max}/E_{\min}$ & $E_g$ & $F_{yy}$ BdG & pred. & $(A_0-F)_{yy}$ & $\Gamma^\kappa_{yy}$ & $\frac12\theta'_y$ & $d^\kappa_\star$ & $d_\star$\\
\midrule
QWZ, $M=-1$ & 0.25 & 3.04 & 40 & 0.118609 & 0.118643 & 0.017081 & 0.017089 & 0.435021 & 0.433457 & 0.488962\\
 & 0.25 & 3.05 & 80 & 0.118507 & 0.118524 & 0.016951 & 0.016955 & 0.433991 & 0.433212 & \\
 & 0.25 & 3.06 & 160 & 0.118456 & 0.118464 & 0.016886 & 0.016889 & 0.433478 & 0.433089 & \\
 & 0.25 & 3.06 & 320 & 0.118430 & 0.118435 & 0.016854 & 0.016855 & 0.433222 & 0.433028 & \\
 & 0.1 & 1.32 & 320 & 0.078193 & 0.078221 & 0.036956 & 0.036968 & 0.476870 & 0.476743 & \\
 & 0.5 & 10.7 & 320 & 0.208265 & 0.20826 & 0.003600 & 0.003601 & 0.36450 & 0.36434 & \\
QWZ, $M=-1.5$ & 0.25 & 3.06 & 320 & 0.112861 & 0.112853 & 0.007435 & 0.007437 & 0.27691 & 0.276808 & 0.343174\\
theta, $d=0.5$ & 0.25 & 3.06 & 160 & 0.119020 & 0.119142 & 0.018358 & 0.018439 & $-0.428676$ & $-0.429357$ & $-0.5$\\
 & 0.25 & 3.06 & 320 & 0.119045 & 0.119106 & 0.01836 & 0.018402 & $-0.428947$ & $-0.429288$ & \\
\bottomrule
\end{tabular}
\end{table}

\subsection*{S9.2 Non-uniform pairing}
Without uniform pairing the same derivation goes through with two changes: the relevant geometry is that of the pairing-deformed band $\tilde u_{\bm k}=D_\Delta^{1/2}u_{\bm k}/|D_\Delta^{1/2}u_{\bm k}|$, $D_\Delta=\mathrm{diag}(\Delta_\alpha)$, and an embedding-independent amplitude term $A_{\rm amp}\succeq0$ appears.  Since $D_\Delta^{1/2}$ is a constant positive matrix, $\tilde u$ has the Chern number of $u$, which gives a floor without uniform pairing: for an exactly flat band at $T=0$, $\sqrt{\det F_{qq}}\ge(|C|/\pi)\min_{\bm k}\Dbar_{\bm k}^2/E_{\bm k}$ with $\Dbar_{\bm k}=\sum_\alpha\Delta_\alpha w_\alpha(\bm k)$.  This is a mean-field statement: without uniform pairing the projected Hubbard model is not a QGN model, and the exactness argument does not apply.  \texttt{bdg\_nonuniform.py} tests it on the flattened QWZ band in the basis $\cos\varphi\,\sigma_z+\sin\varphi\,\sigma_x$ (Table~\ref{tab:nonuni}).  The stiffness residual halves when $E_g$ doubles, the deformed band has $C=-1$ in every case, the phase response follows the deformed natural displacement, and a uniform-pairing formula with the mean gap is off by up to $89\%$.  The BdG omits Hartree shifts, which with unequal orbital densities are orbital dependent; the statement then applies to the Hartree-renormalized band.

\begin{table}[h]
\caption{Non-uniform pairing, flattened QWZ band at $M=-1$ unless noted, $E_g=320|U|$; $48^2$ and $96^2$, and direct minimization and the Schur evaluation, agree to the digits shown.  $\varphi=0$ is the standard basis, $\varphi=\pi/4$ the Letter's.  ``Naive'': uniform-pairing formula with the mean gap.  Numerical.}
\label{tab:nonuni}
\centering
\begin{tabular}{lcccccc}
\toprule
$\varphi/\pi$ & $(\Delta_1,\Delta_2)$ & $F_{xx}$ BdG ; pred. & $F_{yy}$ BdG ; pred. & naive $F_{yy}$ & $\frac12\theta'_y$ ; $\tilde d^\kappa_\star$ & $d_\star$ (undeformed)\\
\midrule
0 & (0.3730, 0.1273) & 0.09606 ; 0.096141 & same & 0.140845 & 0 ; 0 & 0\\
0 ($M=-1.5$) & (0.4163, 0.0840) & 0.07411 ; 0.074172 & same & 0.139981 & 0 ; 0 & 0\\
1/8 & (0.3370, 0.1633) & 0.105337 ; 0.105425 & 0.080354 ; 0.080418 & 0.096157 & 0.409761 ; 0.409771 & 0.422094\\
3/16 & (0.2972, 0.2032) & 0.111194 ; 0.111283 & 0.071332 ; 0.071387 & 0.075189 & 0.47349 ; 0.473497 & 0.477012\\
1/4 & (0.2502, 0.2502) & 0.113408 ; 0.113496 & 0.068097 ; 0.068150 & 0.068150 & 0.48896 ; 0.488962 & 0.488962\\
\bottomrule
\end{tabular}
\end{table}

\subsection*{S9.3 Spin-orbit coupling}
With spin-orbit coupling the isolated band is a Kramers pair of rank two with $C_{\rm tot}=0$.  For on-site singlet pairing each orbital carries one phase shared by both spin components, so the relaxable embeddings are spin locked, $\bm\eta_{\alpha\uparrow}=\bm\eta_{\alpha\downarrow}$, which looks like a splitting defect.  \emph{Proposition.}  If the band is time-reversal symmetric with $\Theta=(\ii s_y\otimes1)\mathsf K$, where $s$ are spin Pauli matrices and $\mathsf K$ is complex conjugation, some natural embedding is spin locked, so $\Sigma_{\rm spin}=0$, and in mean-field theory $\Ds=\frac12\kappa_0G_\star$, with $G_\star$ the unconstrained natural metric of the rank-two Kramers band, which counts both members of each Kramers pair.  \emph{Proof.}  Let $\bm\eta'$ be $\bm\eta$ with the spin labels exchanged.  Then $\Theta P_{\bm\eta}(\bm k)\Theta^{-1}=P_{\bm\eta'}(-\bm k)$, and antiunitary conjugation and $\bm k\to-\bm k$ preserve the metric, so $G_{\bm\eta'}=G_{\bm\eta}$.  The minimizers form an affine set, so if $\bm\eta^\star$ is natural, $(\bm\eta^\star+\bm\eta^{\star\prime})/2$ is natural and spin locked. $\square$  \emph{Stiffness.}  In a basis with $u_n(-\bm k)=\Theta u_n(\bm k)$ the projected singlet pairing at twist $\bm q$ is proportional to $\Delta\,O_{mn}$, $O_{mn}=\langle u_m(\bm k+\bm q)|u_n(\bm k-\bm q)\rangle$, so the flat-band quasiparticle energies are $[\mu^2+\Delta^2\varsigma_n^2]^{1/2}$ with $\varsigma_{1,2}$ the singular values of $O$, and the Nambu sum counts each Kramers pair twice.  Since $\varsigma_1^2+\varsigma_2^2=\Tr[P(\bm k+\bm q)P(\bm k-\bm q)]=2-4g_{ij}q_iq_j+O(q^4)$, with $g$ the metric of the rank-two band, the frozen Hessian is $\frac12\kappa_0G_e$, and Sec.~S4.5 with spin-locked displacements gives $\frac12\kappa_0(G_\star+\Sigma_{\rm spin})=\frac12\kappa_0G_\star$.  \texttt{soc\_bdg.py} tests this on $h=\Xi_\lambda(\bm k)\,\mathrm{diag}(h_\uparrow,h_\downarrow)\,\Xi_\lambda(\bm k)^\dagger$, with $h_\uparrow$ the flattened QWZ band, $h_\downarrow(\bm k)=h_\uparrow(-\bm k)^*$, and $\Xi_\lambda=\exp[\ii\lambda(\cos k_xs_x+\cos k_ys_y)\otimes\tau_z]$, with $\tau_z$ acting on the orbitals, which is time-reversal symmetric, keeps the orbital weights, and mixes the spins oppositely on the two orbitals; it solves the full $8\times8$ Nambu problem.  In units of $\kappa_H$, the relaxed BdG stiffness equals $G_\star/2$ (composite metric per Kramers pair) to the finite-gap residual: at $M=-1$, $\lambda=0.3$, $E_g=160|U|$, $F=\mathrm{diag}(6.15205,4.36325)$ against $\mathrm{diag}(6.15207,4.36326)$, at $\lambda=0.6$, $F=\mathrm{diag}(10.17014,8.38134)$ against $\mathrm{diag}(10.17017,8.38136)$, and at $M=-1.5$, $\lambda=0.6$, $F=\mathrm{diag}(9.5254,8.5952)$ against $\mathrm{diag}(9.5255,8.5952)$, where the digits shown agree at $48^2$ and $96^2$ and between direct minimization and the Schur evaluation, with $\Sigma_{\rm spin}=0$ and the phase response equal to $d_\star$ in every case.  The Chern floor is trivial here because $C_{\rm tot}=0$; a useful floor must come from Wilson-loop bounds, which are stated for the trace \cite{HAPSHB2022}.

\subsection*{S9.4 Real space and composite bands}
For a disordered system on a supercell of $N_c$ cells the band is composite with rank $N_c$, and the weighted Gram theorem, stated for any rank, applies with site-dependent displacements: the natural displacement field, the Loewner-minimal metric, and the site-rigid splitting defect all exist and need only an isolated composite band.  The stiffness identity needs uniform pairing over all sites, that is, equal integrated band density at every site of the supercell \cite{TP2024}, which generic disorder violates.

\section*{S10. The exciton dual}

Take a bilayer whose layers $t$ and $b$ carry the same isolated flat band, with Bloch functions equal up to constant orbital-diagonal phases, no interlayer tunneling, total filling one, and the on-site interlayer interaction $V\sum_{\bm R\alpha}\hat n_{\bm R\alpha t}\hat n_{\bm R\alpha b}$ studied in Ref.~\cite{VGQ2024}.  A particle-hole transformation of the bottom layer, $c_{\bm R\alpha b}\to c^\dagger_{\bm R\alpha b}$, maps (a) the projected bottom band onto $\overline{P(-\bm k)}$, the time-reversed partner of the top band; (b) the interaction onto $-V\sum\bar n_{\bm R\alpha t}\bar n'_{\bm R\alpha}$ plus $V\sum_\alpha\bar w_\alpha\sum_{\bm R}\bar n_{\bm R\alpha t}$, an exact chemical potential under uniform pairing, since $\bar n_{\bm R\alpha b}=\bar w_\alpha-\bar n'_{\bm R\alpha}$ in the projected model; and (c) the counterflow vector potential $(\bm A,-\bm A)$ onto a common one.  Zou \emph{et al.}\ state the map in general \cite{Zou2026}.  The counterflow stiffness is therefore the pair stiffness of our model with $|U|=V$ and $\nu$ the exciton density per cell, and under the same hypotheses (uniform orbital weights, nondegenerate ground state) the identity, the floor, and the exactness argument carry over; the $\eta$-pairing states become the layer-pseudospin ferromagnet.  \texttt{chain\_checks.py}, check (D), confirms the map by exact diagonalization on chains of $6$ and $8$ cells: the four lowest levels of the bilayer, shifted by $Vn_\phi N_p$, equal those of the attractive model to $10^{-14}$ at twists $0$ and $0.13$.  A $\bm k$-dependent relative phase $\mathrm{diag}(1,\e^{\ii k})$ destroys the map (one-pair energy $-0.650\to-0.405$ at $V=1.3$), so the phases must be constant in the convention in which the on-site densities are built.  The remote filled bands must add only orbital-independent Hartree terms, as in the pairing problem.

\section*{S11. Pointwise three-channel decomposition}

With $\lambda_{1,2}$ the eigenvalues of $g$, positivity of $Q$ gives the exact pointwise identity
\begin{equation}
\tr g=|\Om_{xy}|+\bigl(2\sqrt{\det g}-|\Om_{xy}|\bigr)+(\sqrt{\lambda_1}-\sqrt{\lambda_2})^2,
\end{equation}
and after integration $\frac1{2\pi}\int_\BZ\tr g\,\dd^2k=|C|+\mathcal R+\mathcal K+\mathcal S$, with three non-negative channels: curvature-sign reversal $\mathcal R$, failure of pointwise determinant saturation $\mathcal K$ (zero in two-band models), and local anisotropy $\mathcal S$.  The fixed-frame trace defect is $\Ttr=\int(\tr g-|\Om_{xy}|)=2\pi(\mathcal K+\mathcal S)$.  The decomposition holds embedding by embedding, and only $C$ is embedding independent.  \texttt{checks\_identities.py} verifies the pointwise identity on random positive tensors (residual $10^{-14}$), the integrated identity on a random smooth field ($3\times10^{-14}$), and Minkowski's inequality.  \texttt{kappa\_interpolation.py} leaves the elliptic family at its isotropic point ($m^\star=0.029437$) through the non-holomorphic deformation $f_m\to f_m+t(0.3\sin k_x+0.3\ii\sin k_y)$ (the script calls $t$ kappa): at $t=0.05$, $0.2$, and $0.8$, on $256^2$ and $512^2$ grids, the excess of $\sqrt{\det G}$ over $2\pi$ is $0.021\%$, $0.34\%$, and $5.0\%$, carried by the anisotropy channel, while $|C|=2$ and $\mathcal K=0$ throughout; at $t=0.8$ the orbital weights have moved to $0.44/0.56$, so the larger values are statements about band geometry only.

\section*{S12. Reproducibility}

Every number in the Letter and in this Supplemental Material is produced by the scripts listed below; the package README gives the exact commands, and the script \texttt{run\_all.sh} regenerates every output file and figure.  Every output file starts with the command that produced it.  All scripts run with \texttt{python3 -I} and take the scripts directory as their first argument where they import other scripts.  The numbers quoted were obtained with Python 3.13.16, NumPy 2.5.3, SciPy 1.18.1, mpmath 1.4.1, and Matplotlib 3.11.2.

\begingroup
\small
\renewcommand{\arraystretch}{1.15}
\sloppy
\noindent\begin{tabular}{p{0.3\textwidth}p{0.62\textwidth}}
\toprule
script & content\\
\midrule
\raggedright \texttt{checks\_identities.py} & three-channel identity, Minkowski inequality (Sec.~S11)\\
\raggedright \texttt{checks\_gram\_identity.py} & Gram identity and splitting defect on a random four-orbital band (Sec.~S3.7)\\
\raggedright \texttt{check\_qwzx\_rank2.py} & QWZ geometry at two grids; rank-two Gram identity (Sec.~S3.7)\\
\raggedright \texttt{geom.py}, \texttt{wgeom.py} & projector-based geometry for any rank, weighted Gram quantities (library)\\
\raggedright \texttt{qwz\_geom.py}, \texttt{qwzx\_geom.py}, \texttt{qwz\_standard\_basis.py} & QWZ band in both bases (Sec.~S5.6, Table~\ref{tab:qwzgeom})\\
\raggedright \texttt{bdg\_stiffness.py} & self-consistent BdG solver; QWZ family and convergence (Tables~\ref{tab:qwzbdg}, \ref{tab:qwzconv})\\
\raggedright \texttt{theta\_model.py}, \texttt{theta\_inhomogeneity.py}, \texttt{theta\_bdg\_scan.py} & theta band geometry, inhomogeneity, and BdG (Sec.~S5.1)\\
\raggedright \texttt{hbc\_model.py} & $\pi$-flux band (Sec.~S5.2)\\
\raggedright \texttt{liwu\_honeycomb.py} & honeycomb band of Li and Wu (Sec.~S5.3)\\
\raggedright \texttt{ed\_qgn.py}, \texttt{two\_body.py}, \texttt{pair\_mass.py}, \texttt{chain\_checks.py} & exact diagonalization, one-pair curvature, exact pair mass, one-dimensional checks, exciton dual (Secs.~S4, S10)\\
\raggedright \texttt{tbg\_strain.py} and the \texttt{tbg\_*.py} drivers & heterostrained chiral TBG (Sec.~S6)\\
\raggedright \texttt{bdg\_dispersive.py}, \texttt{tdep.py}, \texttt{bdg\_nonuniform.py}, \texttt{soc\_bdg.py}, \texttt{step\_check.py} & robustness (Sec.~S9)\\
\raggedright \texttt{kappa\_interpolation.py}, \texttt{theta\_check.py} & interpolation away from the elliptic family (Sec.~S11); Jacobi-function check (Sec.~S5.4)\\
\raggedright \texttt{make\_fig1\_tradeoff.py}, \texttt{make\_fig2\_theta.py}, \texttt{make\_fig3\_qwz.py}, \texttt{make\_figS1\_elliptic.py}, \texttt{make\_figS2\_tbg.py} & Figs.~1--3 of the Letter and Figs.~S1, S2\\
\bottomrule
\end{tabular}
\par
\endgroup